\documentclass[12pt]{article}
      \def\di{\displaystyle}
      \def\br{{\bf r}}
      \def\bp{{\bf p}}
      \def\bu{{\bf u}}
      \def\bs{{\bf s}}
      \def\R{{\cal R}}
      \def\P{{\cal P}}

      \def\L{{\cal L}}
      \def\N{{\cal N}}
      \def\J{{\cal J}}

\begin{document}

\vskip 5mm
\begin{center}
{\large\bf THE NUCLEAR SCISSORS MODE WITHIN
TWO APPROACHES (Wigner function moments versus RPA)}\\
\vspace*{1cm}
{\large E.B. Balbutsev}\\
\vspace*{0.2cm}
{\it Joint Institute for Nuclear Research, 141980 Dubna, Moscow Region,
Russia}\\
\vspace*{0.5cm}
{\large P. Schuck}\\
\vspace*{0.2cm}
{\it Institut de Physique Nucleaire, Orsay Cedex 91406, France}
\end{center}

      \vspace{3cm}

\begin{abstract}
Two complementary methods to describe the collective motion,
RPA and Wigner function moments method, are compared on an
example of a simple model -- harmonic oscillator with
quadrupole--quadrupole residual interaction. It is
shown that they give identical formulae for eigenfrequencies and
transition probabilities of all collective excitations of the model
including the scissors mode, which here is the subject of our special
attention. The exact relation between the variables of the two
methods and the respective dynamical equations is established.
The normalization factor of the ``synthetic" scissors state and its
overlap with physical states  are calculated analytically. The
orthogonality of the spurious state to all physical states is proved
rigorously.
\end{abstract}

\newpage

\section{Introduction}

The full analysis of the scissors mode in the framework of a solvable
model (harmonic oscillator with quadrupole--quadrupole residual
interaction (HO+QQ)) was
given in \cite{BaSc2}. Several points in the understanding of the
nature of this mode were clarified: for example, its
coexistence with the isovector giant quadrupole resonance (IVGQR),
the decisive role of the Fermi surface deformation, and several
things more.

The Wigner Function Moments (WFM) method was applied to derive
analytical expressions for currents of both coexisting modes
(for the first time), their excitation energies,
magnetic and electric transition probabilities. Our formulae for
energies turned out to be identical with those derived by
Hamamoto and Nazarewicz \cite{Hamam} in the framework of the RPA.
 This fact generated the natural motivation for this work: to
investigate the relation
between formulas for transition probabilities derived by two methods.
More generally we will perform a systematic comparison of the two
approaches. The HO+QQ model is a very convenient ground for
this kind of research, because all results can be obtained
analytically. There is no need to describe the merits
of the RPA -- they are very well known \cite{Ring}.
It is necessary, however, to say a few words about the WFM. Its idea
is based on the virial theorems of Chandrasekhar and Lebovitz
\cite{Chand}. Instead of writing the equations of motion for
microscopic amplitudes of particle--hole excitations (RPA), one writes
the dynamical equations for various multipole phase space moments
of a nucleus. This allows one to achieve a better physical
interpretation of the studied phenomenon without going into its
detailed microscopic structure. The WFM method was successfully
applied to the study of
isoscalar and isovector giant multipole resonances and
low-lying collective modes of rotating and nonrotating nuclei with
various realistic forces \cite{Bal}. The results of WFM
were always very close to similar results obtained with the help of
RPA. In principle, this should be expected, because the basis of both
the methods is the same: Time Dependent Hartree--Fock (TDHF) theory
with its small amplitude approximation. On the other hand, they are
equivalent only in special cases. The detailed analysis of the
interplay of the two methods turns out to be useful also
from a ``practical" point of view: firstly, it allows one to obtain
additional insight into the nature of the scissors mode; secondly,
we find new exact mathematical results for the considered model.

The paper is organized as follows. In Section 2 we remind the
principal points of the WFM formalism and give the summary of the key
results of \cite{BaSc2} obtained by applying this method to the HO+QQ
model. The same model is considered in Section 3 in the frame of RPA:
the formulae for eigenfrequencies, electric and magnetic transition
probabilities of the scissors mode are derived, the ``synthetic"
scissors and spurious state are analyzed, the RPA equations of motion
for transition matrix elements are compared with the WFM equations of
motion for irreducible tensors. The exact relation between the RPA
and WFM variables is established in Section 4. The mutual interplay
of the two methods is discussed in the conclusion. The various
mathematical details are given in Appendices A and B.

\section{The WFM method}

The basis of the method is the Time Dependent Hartree-Fock (TDHF)
equation for the one-body density matrix
$ \rho^{\tau}(\br_1, \br_2, t)=
\langle\br_1|\hat{\rho}^{\tau}(t)|\br_2\rangle $ :
\begin{equation}
\label{TDHF}
 i\hbar \frac{\partial \hat{\rho}^{\tau}}{\partial t}=
\left[ \hat{H}^{\tau} ,\hat{\rho}^{\tau}\right],
\end{equation}
where $ \hat{H}^{\tau} $ is the one-body self-consistent mean field
Hamiltonian depending implicitly on the density matrix and $\tau$ is
an isotopic spin index.
It is convenient to modify equation (\ref{TDHF}) introducing the
Wigner transform of the density matrix
\begin{equation}
\label{f}
 f^{\tau}(\br, \bp, t) = \int d^3s\: \exp(- i\bp
 \cdot \bs/\hbar)\rho^{\tau}(\br+\frac{\bs}{2}, \br-\frac{\bs}{2},t)
\end{equation}
and of the Hamiltonian
\begin{equation}
\label{Hw}
 H_W^{\tau}(\br,\bp)=\int d^3s\: \exp(-i\bp \cdot \bs/\hbar)
(\br+\frac{\bs}{2}\left|\hat{H}^{\tau}\right|\br-\frac{\bs}{2}).
\end{equation}
Using (\ref{f},\ref{Hw}) one arrives \cite{Ring} at

\begin{equation}
\label{fsin}
\frac{\partial f^{\tau}}{\partial t}=
\frac{2}{\hbar}\sin \left\{\frac{\hbar}{2}
\left[(\nabla)^H \cdot (\nabla^p)^f - (\nabla^p)^H \cdot
(\nabla)^f\right]\right\} H_W^{\tau} f^{\tau} ,
\end{equation}
 where the upper index on the bracket stands for the function
on which the operator in these brackets acts.
 It is shown in
\cite{Bal,BaSc}, that by integrating equation (\ref{fsin})
over the phase space $\{\bp,\br\}$ with the weights
$x_{i_1}x_{i_2}\ldots x_{i_k}p_{i_{k+1}}\ldots p_{i_{n-1}}p_{i_n}$,
where $k$ runs from $0$ to $n$, one can obtain a closed finite set
of dynamical equations for Cartesian tensors of the rank $n$.
Taking linear combinations of these equations one is able to represent
them through irreducible tensors, which play the role of collective
variables of the problem. However, it is more convenient to
derive the dynamical equations directly for irreducible tensors using
the technique of tensor products \cite{Varshal}. For this it is
necessary to
rewrite the Wigner function equation (\ref{fsin}) in terms of
cyclic variables
\begin{equation}
\label{sincyclic}
\frac{\partial f^{\tau}}{\partial t}=
\frac{2}{\hbar}\sin \left\{\frac{\hbar}{2}
\sum_{\alpha=-1}^1(-1)^{\alpha}
\left[(\nabla_{-\alpha})^H \cdot (\nabla^p_{\alpha})^f
- (\nabla^p_{-\alpha})^H \cdot (\nabla_{\alpha})^f
\right]\right\} H_W^{\tau} f^{\tau} ,
\end{equation}
 with $$\nabla_{+1}=-\frac{1}{\sqrt 2}(\frac{\partial}{\partial x_1}
+i\frac{\partial}{\partial x_2})~,\quad \nabla_0=\frac{\partial}
{\partial x_3}~,\quad
\nabla_{-1}=\frac{1}{\sqrt 2}(\frac{\partial}{\partial x_1}
-i\frac{\partial}{\partial x_2})~,$$
$$r_{+1}=-\frac{1}{\sqrt 2}(x_1+ix_2)~,\quad r_0=x_3~,\quad
r_{-1}=\frac{1}{\sqrt 2}(x_1-ix_2)$$and the analogous definitions
for $\nabla_{+1}^p~,\quad \nabla_{0}^p~,\quad \nabla_{-1}^p~, $ and
$p_{+1}~,\quad p_{0}~,\quad p_{-1}$.
The required equations are obtained by integrating
(\ref{sincyclic})
with different tensor products of $r_{\alpha}$ and $p_{\alpha}$.
Here we consider the case $n=2$.

\subsection{Model Hamiltonian, Equations of motion}

The microscopic Hamiltonian of the model is
\begin{eqnarray}
\label{Ham}
 H=\sum\limits_{i=1}^A(\frac{\hat\bp_i^2}{2m}+\frac{1}{2}m\omega^2\br_i^2)
+\bar{\kappa}
\sum_{\mu=-2}^{2}(-1)^{\mu}
 \sum\limits_i^Z \sum\limits_j^N
q_{2-\mu}(\br_i)q_{2\mu}(\br_j)
\nonumber\\
+\frac{1}{2}\kappa
\sum_{\mu=-2}^{2}(-1)^{\mu}
\{\sum\limits_{i\neq j}^{Z}
 q_{2-\mu}(\br_i)q_{2\mu}(\br_j)
+\sum\limits_{i\neq j}^{N}
 q_{2-\mu}(\br_i)q_{2\mu}(\br_j)\},
\end{eqnarray}
where the quadrupole operator $q_{2\mu}=\sqrt{16\pi/5}\,r^2Y_{2\mu}$
and $N,Z$ are the numbers of neutrons and protons, respectively.
 The mean field potential for protons (or neutrons) is
\begin{equation}
\label{poten}
V^{\tau}(\br,t)=\frac{1}{2}m\,\omega^2r^2+
\sum_{\mu=-2}^{2}(-1)^{\mu}\tilde Z_{2-\mu}^{\tau}(t)q_{2\mu}(\br),
\end{equation}
where $\tilde Z_{2\mu}^{\rm n}=\kappa Q_{2\mu}^{\rm n}
+\bar{\kappa}Q_{2\mu}^{\rm p}\,,\quad
\tilde Z_{2\mu}^{\rm p}=\kappa Q_{2\mu}^{\rm p}
+\bar{\kappa}Q_{2\mu}^{\rm n}\,$
and the quadrupole moments $Q_{2\mu}^{\tau}(t)$ are defined as
$$ Q_{2\mu}^{\tau}(t)=
\int\! d\{\bp,\br\}
q_{2\mu}(\br)f^{\tau}(\br,\bp,t)$$
 with
 $\int\! d\{\bp,\br\}\equiv
2 (2\pi\hbar)^{-3}\int\! d^3p\,\int\! d^3r$, where the factor 2
appears due to summation over spin degrees of freedom. To simplify
notation we omit spin indices, because we consider spin saturated
system without the spin--orbit interaction.

 Substituting spherical functions by tensor products
$\di r^2Y_{2\mu}=
\sqrt{\frac{15}{8\pi}}r_{2\mu}^2~,$
where
$$r^2_{\lambda\mu}\equiv
\{r\otimes r\}_{\lambda\mu}=\sum_{\sigma,\nu}
C_{1\sigma,1\nu}^{\lambda\mu}r_{\sigma}r_{\nu}$$
and $C_{1\sigma,1\nu}^{\lambda\mu}$ is the Clebsch-Gordan
coefficient,
 one has
\begin{equation}
\label{potenirr}
V^{\tau}=\frac{1}{2}m\,\omega^2r^2
+\sum_{\mu}(-1)^{\mu}Z_{2-\mu}^{\tau}r_{2\mu}^2.
\end{equation}
 Here
$$
Z_{2\mu}^{\rm n}=\chi R_{2\mu}^{\rm n}
+\bar{\chi}R_{2\mu}^{\rm p}\,,\quad
Z_{2\mu}^{\rm p}=\chi R_{2\mu}^{\rm p}
+\bar{\chi}R_{2\mu}^{\rm n}\,,\quad
\chi=6\kappa,\quad\bar\chi=6\bar\kappa,$$
\begin{equation}
\label{Rlmu}
 R_{\lambda\mu}^{\tau}(t)=
\int d\{\bp,\br\}
r_{\lambda\mu}^{2}f^{\tau}(\br,\bp,t).
\end{equation}

 Integration of equation (\ref{sincyclic}) with the weights
$r_{\lambda\mu}^2~,\,
(rp)_{\lambda\mu}\equiv\{r\otimes p\}_{\lambda\mu}$
and $p_{\lambda\mu}^2$
 yields the following set of equations \cite{BaSc2}:
\begin{eqnarray}
\label{quadr}
\frac{d}{dt}R_{\lambda\mu}^{\tau}
-\frac{2}{m}L^{\tau}_{\lambda\mu}&=&0,\quad \lambda=0,2
\nonumber\\
\frac{d}{dt}L^{\tau}_{\lambda\mu}
-\frac{1}{m}P_{\lambda\mu}^{\tau}+
m\,\omega^2R^{\tau}_{\lambda \mu}
-2\sqrt{5}\sum_{j=0}^2\sqrt{2j+1}\{_{2\lambda 1}^{11j}\}
(Z_2^{\tau}R_j^{\tau})_{\lambda \mu}
&=&0,
\quad \lambda=0,1,2
\nonumber\\
\frac{d}{dt}P_{\lambda\mu}^{\tau}
+2m\,\omega^2L^{\tau}_{\lambda \mu}
-4\sqrt{5}\sum_{j=0}^2\sqrt{2j+1}\{_{2\lambda 1}^{11j}\}
(Z_2^{\tau}L^{\tau}_j)_{\lambda \mu}
&=&0,\quad \lambda=0,2
\end{eqnarray}
 where $\{_{2\lambda 1}^{11j}\}$ is the Wigner
$6j$-symbol and the following
notation is introduced
$$P_{\lambda\mu}^{\tau}(t)=
\int\! d\{\bp,\br\}
p_{\lambda\mu}^{2}f^{\tau}(\br,\bp,t),\quad
 L_{\lambda\mu}^{\tau}(t)=
\int\! d\{\bp,\br\}
(rp)_{\lambda\mu}f^{\tau}(\br,\bp,t).$$
By definition
 $q_{2\mu}=\sqrt6r^2_{2\mu},\,
Q_{2\mu}^{\tau}=\sqrt6 R_{2\mu}^{\tau}$,
$R_{00}^{\tau}=-Q_{00}^{\tau}/\sqrt3$
with $Q_{00}^{\tau}=N_{\tau}<r^2>$ being the mean square radius of
neutrons or protons. The tensor $L_{1\nu}^{\tau}$
is connected with angular momentum by the relations
$L_{10}^{\tau}=\frac{i}{\sqrt2}I_3^{\tau},\quad L_{1\pm 1}^{\tau}=
\frac{1}{2}(I_2^{\tau}\mp iI_1^{\tau}).$

  We rewrite equations (\ref{quadr}) in terms
 of the isoscalar and isovector variables
 $R_{\lambda\mu}=R_{\lambda\mu}^{\rm n}+R_{\lambda\mu}^{\rm p},\,
 \bar R_{\lambda\mu}=R_{\lambda\mu}^{\rm n}-R_{\lambda\mu}^{\rm p}$
 (and so on) with the isoscalar $\kappa_0=(\kappa+\bar{\kappa})/2$
 and isovector $\kappa_1=(\kappa-\bar{\kappa})/2$ strength constants.
 There is no problem to solve these equations numerically.
However, we want to simplify the situation as much as possible
to get the results in analytical form giving us a maximum of
insight into the nature of the modes.

1) We consider the problem in small-amplitude approximation.
 Writing all variables as a sum of their
equilibrium value plus a small deviation
$$R_{\lambda\mu}(t)=R_{\lambda\mu}^{eq}+\R_{\lambda\mu}(t),\quad
P_{\lambda\mu}(t)=P_{\lambda\mu}^{eq}+\P_{\lambda\mu}(t),\quad
L_{\lambda\mu}(t)=L_{\lambda\mu}^{eq}+\L_{\lambda\mu}(t),$$
$$\bar R_{\lambda\mu}(t)=\bar R_{\lambda\mu}^{eq}
+\bar \R_{\lambda\mu}(t),\quad
\bar P_{\lambda\mu}(t)=\bar P_{\lambda\mu}^{eq}
+\bar \P_{\lambda\mu}(t),\quad
\bar L_{\lambda\mu}(t)=\bar L_{\lambda\mu}^{eq}
+\bar \L_{\lambda\mu}(t),$$
we linearize the equations of motion in
$\R_{\lambda\mu},\,\P_{\lambda\mu},\,\L_{\lambda\mu}$ and
$\bar \R_{\lambda\mu},\,\bar \P_{\lambda\mu},\,\bar \L_{\lambda\mu}$.

2)We study non--rotating nuclei, i.e. nuclei with
$L_{1\nu}^{eq}=\bar L_{1\nu}^{eq}=0$.

3)Only axially symmetric nuclei with
$R_{2\pm2}^{eq}=R_{2\pm1}^{eq}=\bar R_{2\pm2}^{eq}=
\bar R_{2\pm1}^{eq}=0$ are considered.

4)Finally, we take
\begin{equation}
\label{Apr4}
\bar R_{20}^{eq}=\bar R_{00}^{eq}=0.
\end{equation}
This means that equilibrium deformation and mean square radius of
neutrons are supposed to be equal to that of protons.

Due to the approximation (\ref{Apr4}) the equations for isoscalar and
isovector systems are decoupled. Further, due to the axial symmetry
the angular momentum projection is a good quantum number. As a result,
every set of equations splits into five independent subsets with
$\mu=0,\pm 1,\pm 2.$ The detailed derivation of formulae for
eigenfrequencies and transition probabilities together with all
necessary explanations are given in \cite{BaSc2}. Here we write out
only the final results required for the comparison with respective
results obtained in the framework of RPA.

\subsection{Isoscalar eigenfrequencies}

The isoscalar subset of equations with $\mu=1$ is
\begin{eqnarray}
&&\dot \R_{21}-2\L_{21}/m=0,
\nonumber\\
&&\dot\L_{21}-\P_{21}/m+
\left[m\,\omega^2+ 2\kappa_0
(Q_{20}^{eq}+2Q_{00}^{eq})\right]\R_{21}=0,
\nonumber\\
&&\dot\P_{21}
+2[m\omega^2
+\kappa_0Q_{20}^{eq}]\L_{21}=0,
\nonumber\\
&&\dot \L_{11}=0.
\label{isosca1}
\end{eqnarray}
Imposing the time evolution via $\di{e^{-i\Omega t}}$ for all variables
one transforms (\ref{isosca1}) into a set of algebraic equations.
The eigenfrequencies are found from its characteristic equation
which reads
\begin{equation}
\label{haracis1}
\Omega^2[\Omega^2-4\omega^2-\frac{6\kappa_0}{m}(
Q_{20}^{eq}+\frac{4}{3}Q_{00}^{eq})]=0.
\end{equation}
For $\kappa_0$ we take the self-consistent value
$\di{\kappa_0=-\frac{m\bar{\omega}^2}{4Q_{00}}}$,
where $\di{\bar{\omega}^2=\frac{\omega^2}{1+\frac{2}{3}\delta}}$
(see Appendix~A) with the standard
definition of the deformation parameter
$\di Q_{20}=Q_{00}\frac{4}{3}\,\delta$. Then
\begin{equation}
\label{haracis2}
\Omega^2[\Omega^2-2\bar{\omega}^2(1+\delta/3)]=0.
\end{equation}
The nontrivial solution of this equation gives the frequency of the
$\mu=1$ branch of the isoscalar GQR
\begin{equation}
\label{omegis}
\Omega^2=\Omega_{is}^2=2\bar{\omega}^2(1+\delta/3).
\end{equation}
Taking into account the relation (A.7) we find that this
result coincides with that of \cite{Suzuki}.
The trivial solution $\Omega=\Omega_0=0$ is characteristic of
nonvibrational mode corresponding to the obvious integral of motion
$\L_{11}=const$ responsible for the rotational degree of freedom.
This is usually called the `spurious' or `Goldstone' mode.

\subsection{Isovector eigenfrequencies}

The information about the scissors mode is contained in the
subset of isovector equations
with $\mu=1$
\begin{eqnarray}
\label{scis}
&&\dot{\bar\R}_{21}-2\bar\L_{21}/m=0,
\nonumber\\
&&\dot{\bar\L}_{21}-\bar\P_{21}/m
+\left[m\,\omega^2+\kappa Q_{20}^{eq}
+4\kappa_1Q_{00}^{eq}\right]\bar\R_{21}=0,
\nonumber\\
&&\dot{\bar\P}_{21}
+2[m\omega^2+\kappa_0Q_{20}^{eq}]\bar\L_{21}
-6\kappa_0Q_{20}^{eq}\,\bar\L_{11}=0,
\nonumber\\
&&\dot{\bar\L}_{11}
+3\bar{\kappa}Q_{20}^{eq}\bar\R_{21}=0.
\end{eqnarray}
Imposing the time evolution via $\di{e^{-i\Omega t}}$
one transforms (\ref{scis}) into a set of algebraic equations.
Again the eigenfrequencies are found from the characteristic equation
which reads
\begin{equation}
\label{harac1}
\Omega^4-\Omega^2[4\omega^2+\frac{8}{m}\kappa_1Q_{00}^{eq}
+\frac{2}{m}(\kappa_1+2\kappa_0)Q_{20}^{eq}]
+\frac{36}{m^2}(\kappa_0-\kappa_1)\kappa_0(Q_{20}^{eq})^2=0.
\end{equation}
Supposing, as usual, the isovector constant $\kappa_1$ to be
proportional to the isoscalar one, $\kappa_1=\alpha\kappa_0$, and
taking the self-consistent value for $\kappa_0$,
we finally obtain
\begin{equation}
\label{harac2}
\Omega^4-2\Omega^2\bar{\omega}^2(2-\alpha)(1+\delta/3)
+4\bar{\omega}^4(1-\alpha)\delta^2=0.
\end{equation}
The solutions of this equation are
\begin{equation}
\label{Omeg2}
\Omega^2_{\pm}=\bar{\omega}^2(2-\alpha)(1+\delta/3)
\pm \sqrt{\bar{\omega}^4(2-\alpha)^2(1+\delta/3)^2
-4\bar{\omega}^4(1-\alpha)\delta^2}.
\end{equation}
The high-lying solution $\Omega_+$ gives the frequency $\Omega_{iv}$
of the $\mu=1$ branch of the isovector GQR.
The low-lying solution $\Omega_-$ gives the frequency $\Omega_{sc}$
of the scissors mode.

 We adjust $\alpha$ from the fact that the IVGQR is
experimentally known to lie practically at twice the energy of the
isoscalar GQR. In our model the experimental situation is satisfied
by $\alpha=-2$. Then
\begin{eqnarray}
\Omega^2_{iv}=4\bar\omega^2
\left(1+\frac{\delta}{3}+\sqrt{(1+\frac{\delta}{3})^2-
 \frac{3}{4}\delta^2}\,\right),
\quad
\Omega^2_{sc}=4\bar\omega^2
\left(1+\frac{\delta}{3}-\sqrt{(1+\frac{\delta}{3})^2-
\frac{3}{4}\delta^2}\,\right).
\label{Omeg2fin}
\end{eqnarray}

\subsection{Linear response and transition probabilities}

A direct way of calculating the reduced transition probabilities
is provided by the theory of the linear response of a system to a
weak external field
$$\hat F(t)=\hat F\,{\rm exp}(-i\Omega t)+
\hat F^{\dagger}\,{\rm exp}(i\Omega t),$$
where $\hat F=\sum_{s=1}^A\hat f_s$ is a one-body operator.
A convenient form of the response theory is e.g. given by Lane
\cite{Lan} (see also section 4).
The matrix elements of the operator $\hat F$ obey the relation
\begin{equation}
\label{Fmatel}
|<\nu|\hat F|0>|^2=
\hbar\lim_{\Omega\to\Omega_{\nu}}(\Omega-\Omega_{\nu})
\overline{<\psi|\hat F|\psi>\exp(-i\Omega t)},
\end{equation}
where $|0>$ and $|\nu>$ are the stationary wave functions of the
ground and unperturbed excited states; $\psi$ is the perturbed
wavefunction of the ground state, $\Omega_{\nu}=(E_{\nu}-E_0)/\hbar$
are the
normal frequencies, the bar means averaging over a time interval much
larger than $1/\Omega$, $\Omega$ being the frequency of the external
field $\hat F(t)$.

{\bf Magnetic excitations}
\begin{equation}
\label{Omagn}
\hat F=\hat F_{1\mu}^{\rm p}=\sum_{s=1}^Z\hat f_{1\mu}(s), \quad
\hat f_{1\mu}=-i\nabla
(rY_{1\mu})\cdot[\br\times\nabla]\mu_N=\gamma(r\hat p)_{1\mu},\quad
\mu_N=\frac{e\hbar}{2mc}.
\end{equation}
$$<\psi|\hat F_{1\mu}^{\rm p}|\psi>=\gamma L_{1\mu}^{\rm p}=
\frac{\gamma}{2}(L_{1\mu}-\bar L_{1\mu})=
\frac{\gamma}{2}(\L_{1\mu}-\bar \L_{1\mu}),\quad
\gamma=-i\frac{e}{2mc}\sqrt{\frac{3}{2\pi}}.$$
\begin{eqnarray}
\label{scimat}
B(M1)_{sc}=2|<sc|\hat F_{11}^{\rm p}|0>|^2=
\frac{1-\alpha}{4\pi}\frac{m\bar\omega^2}{\hbar}
Q_{00}\delta^2\frac{\Omega_{sc}^2-2(1+\delta/3)\bar\omega^2}
{\Omega_{sc}(\Omega^2_{sc}-\Omega_{iv}^2)}\,\mu_N^2,
\end{eqnarray}
\begin{eqnarray}
\label{M1iv}
B(M1)_{iv}=2|<iv|\hat F_{11}^{\rm p}|0>|^2
=\frac{1-\alpha}{4\pi}\frac{m\bar\omega^2}{\hbar}Q_{00}
\delta^2\frac{\Omega_{iv}^2-2(1+\delta/3)\bar\omega^2}
{\Omega_{iv}(\Omega^2_{iv}-\Omega_{sc}^2)}\,\mu_N^2.
\end{eqnarray}
These two formulae can be joined into one expression
by the simple transformation of the denominators. Really,
we have from (\ref{Omeg2})
\begin{eqnarray}
\pm(\Omega^2_{iv}-\Omega_{sc}^2)&=&\pm(\Omega^2_{+}-\Omega_{-}^2)=
\pm 2 \sqrt{\bar{\omega}^4(2-\alpha)^2(1+\delta/3)^2
-4\bar{\omega}^4(1-\alpha)\delta^2}
\nonumber\\
&=&2\Omega^2_{\pm}-2\bar{\omega}^2(2-\alpha)(1+\delta/3)=
2\Omega^2_{\pm}-(2-\alpha)(\omega_x^2+\omega_z^2).
\label{Deno}
\end{eqnarray}
Using these relations in formulae (\ref{scimat}) and (\ref{M1iv}), we
obtain the expression for the $B(M1)$ values valid for both excitations
\begin{eqnarray}
\label{BM1gen}
B(M1)_{\nu}=2|<\nu|\hat F_{11}^{\rm p}|0>|^2
=\frac{1-\alpha}{8\pi}\frac{m\bar\omega^2}{\hbar}Q_{00}\delta^2
\frac{\Omega_{\nu}^2-2(1+\delta/3)\bar\omega^2}
{\Omega_{\nu}[\Omega^2_{\nu}-\bar{\omega}^2(2-\alpha)(1+\delta/3)]}
\,\mu_N^2.
\end{eqnarray}

{\bf Electric excitations}

\begin{equation}
\label{Oelec}
\hat F=\hat F_{2\mu}^{\rm p}=\sum_{s=1}^Z\hat f_{2\mu}(s), \quad
\hat f_{2\mu}=e\,r^2Y_{2\mu}=\beta r^2_{2\mu},\quad
\beta=e\sqrt{\frac{15}{8\pi}}.
\end{equation}
$$<\psi|\hat F_{2\mu}^{\rm p}|\psi>=\beta R_{2\mu}^{\rm p}=
\frac{1}{2}\beta (R_{2\mu}-\bar R_{2\mu}).$$

\begin{eqnarray}
\label{E2sc}
B(E2)_{sc}=2|<sc|\hat F_{21}^{\rm p}|0>|^2
=\frac{e^2\hbar}{m}\frac{5}{8\pi}Q_{00}
\frac{(1+\delta/3)\Omega_{sc}^2-2(\bar\omega\delta)^2}
{\Omega_{sc}(\Omega^2_{sc}-\Omega_{iv}^2)}.
\end{eqnarray}

\begin{eqnarray}
\label{E2iv}
B(E2)_{iv}=2|<iv|\hat F_{21}^{\rm p}|0>|^2
=\frac{e^2\hbar}{m}\frac{5}{8\pi}Q_{00}
\frac{(1+\delta/3)\Omega_{iv}^2-2(\bar\omega\delta)^2}
{\Omega_{iv}(\Omega^2_{iv}-\Omega_{sc}^2)}.
\end{eqnarray}

\begin{eqnarray}
\label{E2is}
B(E2)_{is}=2|<is|\hat F_{21}^{\rm p}|0>|^2
=\frac{e^2\hbar}{m}\frac{5}{8\pi}Q_{00}
[(1+\delta/3)\Omega_{is}^2
-2(\bar\omega\delta)^2]/[\Omega_{is}]^3.
\end{eqnarray}
Using relations (\ref{Deno}) in formulae (\ref{E2sc}) and
(\ref{E2iv}) we
obtain the expression for the $B(E2)$ values valid for all three
excitations
\begin{eqnarray}
\label{BE2gen}
B(E2)_{\nu}=2|<\nu|\hat F_{21}^{\rm p}|0>|^2
=\frac{e^2\hbar}{m}\frac{5}{16\pi}Q_{00}
\frac{(1+\delta/3)\Omega_{\nu}^2-2(\bar\omega\delta)^2}
{\Omega_{\nu}[\Omega^2_{\nu}-\bar{\omega}^2(2-\alpha)(1+\delta/3)]}.
\end{eqnarray}
The isoscalar value (\ref{E2is}) is obtained by assuming $\alpha=1.$

\section{Random Phase Approximation (RPA)}

 Standard RPA equations in the notation of \cite{Ring} are
\begin{eqnarray}
\label{RPA}
\sum_{n,j}\left\{\left[\delta_{ij}\delta_{mn}(\epsilon_m-
\epsilon_i)+\bar v_{mjin}\right]X_{nj}+\bar v_{mnij}Y_{nj}\right\}
=\hbar\Omega X_{mi},
\nonumber\\
\sum_{n,j}\left\{\bar v_{ijmn}X_{nj}+
\left[\delta_{ij}\delta_{mn}(\epsilon_m-
\epsilon_i)+\bar v_{inmj}\right]Y_{nj}\right\}
=-\hbar\Omega Y_{mi}.
\end{eqnarray}
According to the definition of the schematic model by \cite{Ring},
the matrix elements of the residual interaction corresponding to the
Hamiltonian (\ref{Ham}) are
$$\bar v_{mjin}=
\kappa_{\tau\tau'}D_{im}^{\tau*}D_{jn}^{\tau'}$$
with $D_{im}\equiv<i|q_{21}|m>$ and $\kappa_{\rm nn}
=\kappa_{\rm pp}=\kappa,\quad \kappa_{\rm np}=\bar\kappa$.
This interaction distinguishes between protons and neutrons, so we have
to introduce the isospin indices $\tau,\,\tau'$
 into the set of RPA equations
(\ref{RPA}):
\begin{eqnarray}
\label{DD}
(\epsilon_m^{\tau}-\epsilon_i^{\tau})X_{mi}^{\tau}+
\sum_{n,j,\tau'}\kappa_{\tau\tau'}D_{im}^{\tau*}D_{jn}^{\tau'}
X_{nj}^{\tau'}+
\sum_{n,j,\tau'}\kappa_{\tau\tau'}D_{im}^{\tau*}D_{nj}^{\tau'}
Y_{nj}^{\tau'}
=\hbar\Omega X_{mi}^{\tau},
\nonumber\\
\sum_{n,j,\tau'}\kappa_{\tau\tau'}D_{mi}^{\tau*}D_{jn}^{\tau'}
X_{nj}^{\tau'}+
(\epsilon_m^{\tau}-\epsilon_i^{\tau})Y_{mi}^{\tau}+
\sum_{n,j,\tau'}\kappa_{\tau\tau'}D_{mi}^{\tau*}D_{nj}^{\tau'}
Y_{nj}^{\tau'}
=-\hbar\Omega Y_{mi}^{\tau}.
\end{eqnarray}
 The solution is
\begin{equation}
X_{mi}^{\tau}=
\frac{D_{im}^{\tau*}}{\hbar\Omega-\epsilon_{mi}^{\tau}}K^{\tau},
\quad
Y_{mi}^{\tau}=
-\frac{D_{mi}^{\tau*}}{\hbar\Omega+\epsilon_{mi}^{\tau}}K^{\tau}
\label{XY}
\end{equation}
with $\epsilon_{mi}^{\tau}=\epsilon_m^{\tau}-\epsilon_i^{\tau}$ and
$K^{\tau}=\sum_{\tau'}\kappa_{\tau\tau'}C^{\tau'}.$

The constant $C^{\tau}$ is defined as
$C^{\tau}=\sum_{n,j}(D_{jn}^{\tau}X_{nj}^{\tau}
+D_{nj}^{\tau}Y_{nj}^{\tau}).$
Using here the expressions for $X_{nj}^{\tau}$ and
$Y_{nj}^{\tau}$ given above,
one derives the useful relation
\begin{equation}
C^{\tau}=2S^{\tau}K^{\tau}
=2S^{\tau}\sum_{\tau'}\kappa_{\tau\tau'}C^{\tau'},
\label{CSK}
\end{equation}
where the following notation is introduced:
\begin{equation}
S^{\tau}=\sum_{mi}|D_{mi}^{\tau}|^2\frac{\epsilon_{mi}^{\tau}}
{E^2-(\epsilon_{mi}^{\tau})^2}
\label{S}
\end{equation}
with $E=\hbar\Omega$.
Let us write out the relation (\ref{CSK}) in detail
\begin{eqnarray}
C^{\rm n}-2S^{\rm n}(\kappa C^{\rm n}+\bar\kappa C^{\rm p})=0,
\nonumber\\
C^{\rm p}-2S^{\rm p}(\bar\kappa C^{\rm n}+\kappa C^{\rm p})=0.
\label{Cnp}
\end{eqnarray}
The condition for existence of a nontrivial solution of this set of
equations gives the secular equation
\begin{equation}
(1-2S^{\rm n}\kappa)(1-2S^{\rm p}\kappa)
-4S^{\rm n}S^{\rm p}\bar\kappa^2=0.
\label{Secul1}
\end{equation}
Making obvious linear combinations of the two equations in
(\ref{Cnp}), we write them in terms of isoscalar and
isovector constants $C=C^{\rm n}+C^{\rm p},\,\bar C=C^{\rm n}-C^{\rm p}$
\begin{eqnarray}
C-2(S^{\rm n}+S^{\rm p})\kappa_0 C
-2(S^{\rm n}-S^{\rm p})\kappa_1 \bar C=0,
\nonumber\\
\bar C-2(S^{\rm n}-S^{\rm p})\kappa_0 C
-2(S^{\rm n}+S^{\rm p})\kappa_1 \bar C=0.
\label{Cnpiso}
\end{eqnarray}
 Approximation (\ref{Apr4}) allows us to decouple the equations for
isoscalar and isovector constants. Really, in this case
$S^{\rm n}=S^{\rm p}\equiv S/2$;
hence, we obtain two secular equations
\begin{equation}
1-2S\kappa_0=0,
 \quad \mbox{or}\quad 1-S\kappa=S\bar\kappa
\label{Secis}
\end{equation}
in the isoscalar case and
\begin{equation}
1-2S\kappa_1=0,
 \quad \mbox{or}\quad 1-S\kappa=-S\bar\kappa
\label{Seciv}
\end{equation}
in the isovector one, the difference between them being in the
strength constants only. Having in mind the relation
$\kappa_1=\alpha \kappa_0$, we come to the conclusion
that it is sufficient to analyze the isovector case only -- the
results for isoscalar one are obtained by assuming $\alpha=1$.

\subsection{Eigenfrequencies}

The detailed expression for the isovector secular equation is
\begin{equation}
\frac{1}{2\kappa_1}=
\sum_{mi}|D_{mi}|^2\frac{\epsilon_{mi}}
{E^2-\epsilon_{mi}^2}.
\label{Secular}
\end{equation}
The operator $D$ has only two types of
nonzero matrix elements $D_{mi}$ in the deformed oscillator basis.
Matrix elements of the first type couple states of the same major
shell. All corresponding transition energies are degenerate:
$\epsilon_m-\epsilon_i=\hbar(\omega_x-\omega_z)\equiv\epsilon_0$.
Matrix elements of the second type couple states of the different
major shells with $\Delta N=2$. All corresponding transition energies
are degenerate too:
$\epsilon_m-\epsilon_i=\hbar(\omega_x+\omega_z)\equiv\epsilon_2$.
Therefore, the secular equation can be rewritten as
\begin{equation}
\frac{1}{2\kappa_1}=\frac{\epsilon_0D_0}{E^2-\epsilon_0^2}+
\frac{\epsilon_2D_2}{E^2-\epsilon_2^2}.
\label{Secul}
\end{equation}
The sums $\di D_0=\sum_{mi(\Delta N=0)}|D_{mi}|^2$ and
$\di D_2=\sum_{mi(\Delta N=2)}|D_{mi}|^2$ can be calculated
analytically (see Appendix B):
\begin{equation}
\label{sums}
D_0=\frac{Q_{00}}{m\bar\omega^2}\epsilon_0,
\quad
D_2=\frac{Q_{00}}{m\bar\omega^2}\epsilon_2.
\end{equation}
Let us transform the secular equation (\ref{Secul}) in polynomial form
$$E^4-E^2[(\epsilon_0^2+\epsilon_2^2)+2\kappa_1(\epsilon_0D_0
+\epsilon_2D_2)]+[\epsilon_0^2\epsilon_2^2
+2\kappa_1\epsilon_0\epsilon_2(\epsilon_0D_2+\epsilon_2D_0)]=0.$$
Using here the expressions (\ref{sums}) for $D_0,\,D_2$ and the
self-consistent value of the strength constant (A.3), we find
$$E^4-E^2(1-\alpha/2)(\epsilon_0^2+\epsilon_2^2)
+(1-\alpha)\epsilon_0^2\epsilon_2^2=0,$$ or
\begin{equation}
\label{Secpol}
\Omega^4-\Omega^2(2-\alpha)\omega_+^2+(1-\alpha)\omega_-^4=0,
\end{equation}
with the notation $\omega_+^2=\omega_x^2+\omega_z^2$ and
$\omega_-^4=(\omega_x^2-\omega_z^2)^2$.
This result coincides with that of \cite{Hamam}. By a trivial
rearrangement of the terms in (\ref{Secpol}) one obtains the useful
relation
\begin{equation}
\label{Relat}
\Omega^2(\Omega^2-\omega_+^2)=
(1-\alpha)(\Omega^2\omega_+^2-\omega_-^4).
\end{equation}
Inserting expressions (A.3) for $\omega_x^2,\,\omega_z^2$
into (\ref{Secpol}), we find
$\omega_+^2=2\bar\omega^2(1+\delta/3),\,
\omega_-^4=4\delta^2\bar\omega^4$
and reproduce formula (\ref{harac2}) for the isovector case
$$\Omega^4-2\Omega^2\bar{\omega}^2(2-\alpha)(1+\delta/3)
+4\bar{\omega}^4(1-\alpha)\delta^2=0.$$
Taking here $\alpha=1$ we reproduce formula (\ref{haracis2}) for
the isoscalar case
$$\Omega^4-2\Omega^2\bar{\omega}^2(1+\delta/3)=0.$$

\subsection{B(E2)-factors}

 According to \cite{Ring}, the transition probability for the one-body
operator $\di{\hat F=\sum_{s=1}^A\hat f_s}$
is calculated by means of the formulae
\begin{equation}
\label{matelem}
<0|\hat F^{\tau}|\nu>=\sum_{mi}(f_{im}^{\tau}X_{mi}^{\tau,\nu}
+f_{mi}^{\tau}Y_{mi}^{\tau,\nu}),\quad
<\nu|\hat F^{\tau}|0>=\sum_{mi}(f_{mi}^{\tau}X_{mi}^{\tau,\nu}
+f_{im}^{\tau}Y_{mi}^{\tau,\nu}).
\end{equation}
Quadrupole excitations are described by the operator (\ref{Oelec})
with $\hat f_{2\mu}=er^2Y_{2\mu}=\tilde eD$, where
$\tilde e=e\sqrt{\frac{5}{16\pi}}$.
The expressions for $X_{mi}^{\tau},\,Y_{mi}^{\tau}$ are given by
formulae (\ref{XY}). Combining these results we get
\begin{equation}
<0|\hat F_{21}^{\rm p}|\nu>=2\tilde e
K_{\nu}^{\rm p}\sum_{mi}|D_{mi}^{\rm p}|^2
\frac{\epsilon_{mi}^{\rm p}}{E_{\nu}^2-(\epsilon_{mi}^{\rm p})^2}=
2\tilde eK_{\nu}^{\rm p}S_{\nu}^{\rm p}=\tilde eC_{\nu}^{\rm p}.
\label{probab}
\end{equation}
The constant $C_{\nu}^{\rm p}$
is determined by the normalization condition
$$\delta_{\nu,\nu'}=\sum_{mi,\tau}(X_{mi}^{\tau,\nu*}X_{mi}^{\tau,\nu'}
-Y_{mi}^{\tau,\nu*}Y_{mi}^{\tau,\nu'}),$$
that gives
\begin{equation}
\frac{1}{(C_{\nu}^{\rm p})^2}=E_{\nu}\sum_{mi}\left[
\frac{|D_{mi}^{\rm p}|^2}{(S_{\nu}^{\rm p})^2}
\frac{\epsilon_{mi}^{\rm p}}{[E_{\nu}^2-(\epsilon_{mi}^{\rm p})^2]^2}
+\frac{(C_{\nu}^{\rm n})^2}{(C_{\nu}^{\rm p})^2}
\frac{|D_{mi}^{\rm n}|^2}{(S_{\nu}^{\rm n})^2}
\frac{\epsilon_{mi}^{\rm n}}{[E_{\nu}^2-(\epsilon_{mi}^{\rm n})^2]^2}
\right].
\label{norma}
\end{equation}
The ratio $C^{\rm n}/C^{\rm p}$ is determined by any of the equations
(\ref{Cnp}):
\begin{equation}
\frac{C^{\rm n}}{C^{\rm p}}
=\frac{1-2S^{\rm p}\kappa}{2S^{\rm p}\bar\kappa}
=\frac{2S^{\rm n}\bar\kappa}{1-2S^{\rm n}\kappa}.
\label{Cn/Cp}
\end{equation}
Formula (\ref{norma}) is considerably simplified by the approximation
(\ref{Apr4}), when $S^{\rm p}=S^{\rm n}\equiv S/2,\,
\epsilon_{mi}^{\rm p}=\epsilon_{mi}^{\rm n},\,
D_{mi}^{\rm p}=D_{mi}^{\rm n}$. Applying the second forms of
formulae (\ref{Secis},\,\ref{Seciv}) it is easy to find that in this
case $C^{\rm n}/C^{\rm p}=\pm 1$. As a result,
the final expression for $B(E2)$ value is
\begin{equation}
B(E2)_{\nu}=2|<0|\hat F_{21}^{\rm p}|\nu>|^2
=2\tilde e^2\left(16E_{\nu}\kappa_1^2\sum_{mi}|D_{mi}|^2
\frac{\epsilon_{mi}}{(E_{\nu}^2-\epsilon_{mi}^2)^2}\right)^{-1}.
\label{BE2}
\end{equation}
With the help of formulae (\ref{sums}) this expression can be
transformed into
\begin{eqnarray}
B(E2)_{\nu}&=&\frac{5}{8\pi}\frac{e^2Q_{00}}{m\bar\omega^2\alpha^2E_{\nu}}
\left[\frac{\epsilon_0^2}{(E_{\nu}^2-\epsilon_0^2)^2}
     +\frac{\epsilon_2^2}{(E_{\nu}^2-\epsilon_2^2)^2}\right]^{-1}
\nonumber\\
&=&\frac{5}{8\pi} \frac{e^2Q_{00}}{m\bar\omega^2\alpha^2E_{\nu}}
\frac{(E_{\nu}^2-\epsilon_0^2)^2(E_{\nu}^2-\epsilon_2^2)^2}
{(E_{\nu}^2-\epsilon_2^2)^2\epsilon_0^2
+(E^2-\epsilon_0^2)^2\epsilon_2^2}
\nonumber\\
&=&\frac{5}{16\pi}\frac{e^2\hbar Q_{00}}{m\bar\omega^2\Omega_{\nu}}
\frac{(\Omega_{\nu}^2\omega_+^2 -\omega_-^4)^2}
{\Omega_{\nu}^4\omega_+^2 -2\Omega_{\nu}^2\omega_-^4
+\omega_+^2\omega_-^4}.
\label{BE2rpa}
\end{eqnarray}
At first sight, this expression has nothing in common with
(\ref{BE2gen}).
Nevertheless, it can be shown that they are identical. To this end,
we analyze carefully the denominator of the last expression in
(\ref{BE2rpa}). Summing it with the secular equation (\ref{Secpol})
(multiplied by $\omega_+^2$), which
obviously does not change its value, we find after elementary
combinations
\begin{eqnarray}
{\rm Denom}&=&\Omega_{\nu}^4\omega_+^2 -2\Omega_{\nu}^2\omega_-^4
+\omega_+^2\omega_-^4
+\omega_+^2[\Omega_{\nu}^4-\Omega_{\nu}^2(2-\alpha)\omega_+^2
+(1-\alpha)\omega_-^4]
\nonumber\\
&=&\omega_+^2\Omega_{\nu}^2[2\Omega_{\nu}^2-(2-\alpha)\omega_+^2]
-\omega_-^4[2\Omega_{\nu}^2-(2-\alpha)\omega_+^2]
\nonumber\\
&=&(\Omega_{\nu}^2\omega_+^2-\omega_-^4)
[2\Omega_{\nu}^2-(2-\alpha)\omega_+^2].
\label{denom}
\end{eqnarray}
This result allows us to write the final expression
\begin{eqnarray}
B(E2)_{\nu}=
\frac{5}{16\pi}\frac{e^2\hbar}{m\bar\omega^2}Q_{00}
\frac{\Omega_{\nu}^2\omega_+^2 -\omega_-^4}
{\Omega_{\nu}[2\Omega_{\nu}^2-(2-\alpha)\omega_+^2]},
\label{BE2fin}
\end{eqnarray}
which coincides with (\ref{BE2gen}) (we recall that
$\omega_+^2=2\bar\omega^2(1+\delta/3),\,
\omega_-^4=4\delta^2\bar\omega^4$).
By the simple transformations
this formula is reduced to the result of Hamamoto and Nazarewicz
\cite{Hamam} (taking into account, that they published it without the
constant factor
$\di{\frac{5}{32\pi}\frac{e^2\hbar}{m\omega_0}Q_{00}^0}$).

\subsection{B(M1)-factors}

In accordance with formulae (\ref{Omagn}), (\ref{matelem}),
(\ref{XY}) the magnetic transition matrix element is
\begin{equation}
<0|\hat F^{\rm p}_{11}|\nu>=K_{\nu}^{\rm p}\sum_{mi}\left[
\frac{(\hat f^{\rm p}_{11})_{im}D_{im}^{\rm p*}}
{E_{\nu}-\epsilon_{mi}^{\rm p}}-
\frac{(\hat f^{\rm p}_{11})_{mi}D_{mi}^{\rm p*}}
{E_{\nu}+\epsilon_{mi}^{\rm p}}
\right].
\label{Mprob}
\end{equation}
As it is shown in Appendix B, the matrix element
$(f^{\rm p}_{11})_{im}$ is proportional to $D_{im}^{\rm p}$ (formula
(B.16). So, expression (\ref{Mprob}) is reduced to
\begin{eqnarray}
<0|\hat F^{\rm p}_{11}|\nu>&=&-K_{\nu}^{\rm p}
\frac{\tilde e\hbar}{2c\sqrt5}
(\omega_x^2-\omega_z^2)^{\rm p}\sum_{mi}\left[
\frac{D^{\rm p}_{im}D_{im}^{\rm p*}}
{\epsilon_{im}^{\rm p}(E_{\nu}-\epsilon_{mi}^{\rm p})}-
\frac{D^{\rm p}_{mi}D_{mi}^{\rm p*}}
{\epsilon_{mi}^{\rm p}(E_{\nu}+\epsilon_{mi}^{\rm p})}\right]
\nonumber\\
&=&K_{\nu}^{\rm p}\frac{\tilde e\hbar}{c\sqrt5}
(\omega_x^2-\omega_z^2)^{\rm p}E_{\nu}\sum_{mi}
\frac{|D^{\rm p}_{mi}|^2}
{\epsilon_{mi}^{\rm p}[E_{\nu}^2-(\epsilon_{mi}^{\rm p})^2]}.
\label{Magnet}
\end{eqnarray}
With the help of approximation (\ref{Apr4}) and the expressions
(\ref{sums}) for $D_0,\,D_2$ we find
\begin{eqnarray}
<0|\hat F^{\rm p}_{11}|\nu>
&=&\frac{C_{\nu}^{\rm p}}{2S_{\nu}^{\rm p}}
\frac{\tilde e\hbar}{c\sqrt5}(\omega_x^2-\omega_z^2)
\frac{Q_{00}}{2m\bar\omega^2}
(\frac{E_{\nu}}{E_{\nu}^2-\epsilon_0^2}
+\frac{E_{\nu}}{E_{\nu}^2-\epsilon_2^2})
\nonumber\\
&=&-2\kappa_1C_{\nu}^{\rm p}\frac{\tilde e}{c\sqrt5}
(\omega_x^2-\omega_z^2)\frac{Q_{00}}{m\bar\omega^2}
\frac{\Omega_{\nu}(\Omega_{\nu}^2-\omega_+^2)}
{\alpha(\Omega_{\nu}^2\omega_+^2-\omega_-^4)}
\nonumber\\
&=&\frac{C_{\nu}^{\rm p}}{2}\frac{\tilde e}{c\sqrt5}
(\omega_x^2-\omega_z^2)\frac{1-\alpha}{\Omega_{\nu}}.
\label{Magnet1}
\end{eqnarray}
Relation (\ref{Relat}) and the self-consistent value of the
strength constant $\kappa_1=\alpha\kappa_0$ were used in the last
step. For the magnetic transition probability we have
\begin{eqnarray}
B(M1)_{\nu}=2|<0|\hat F^{\rm p}_{11}|\nu>|^2
=2\frac{(C_{\nu}^{\rm p})^2}{4}\frac{\tilde e^2}{5c^2}
\omega_-^4\frac{(1-\alpha)^2}{\Omega_{\nu}^2}
=\frac{\omega_-^4}{20c^2}
\frac{(1-\alpha)^2}{\Omega_{\nu}^2}B(E2)_{\nu}.
\label{BM1rpa}
\end{eqnarray}
This relation between $B(M1)$ and $B(E2)$ was also found (up to the
factor $1/(20c^2)$) by Hamamoto and Nazarewicz \cite{Hamam}.
Substituting expression (\ref{BE2fin}) for $B(E2)$ into (\ref{BM1rpa})
we reproduce (with the help of relation (\ref{Relat})) formula
(\ref{BM1gen}).

\subsection{``Synthetic" scissors and spurious state}

 The nature of collective excitations calculated with the method of
Wigner function moments is quite easily revealed analyzing the
roles of collective variables describing the
phenomenon. The solution of this problem in the RPA approach is not so
obvious. That is why the nature of the low-lying states has often been
established by considering overlaps of these states with the "pure
scissors state" \cite{Hilt86,Dieper} or "synthetic state" \cite{Hamam}
produced by the action of the scissors operator
$$\hat S_x=\N^{-1}(<{I_x^{\rm n}}^2>\hat I_x^{\rm p}-
<{I_x^{\rm p}}^2>\hat I_x^{\rm n})$$
on the ground state
$$|Syn>=\hat S_x|0>.$$
In the considered model the overlap of the ``synthetic" state with
the real scissors mode (and with IVGQR) can be calculated
analytically. Surprisingly, it was not done until now. Let us at first
modify the definition of the ``synthetic" state.
Due to axial symmetry one can use the $\hat I_y^{\tau}$ component
instead of $\hat I_x^{\tau}$, or any of their linear combinations,
for example, the $\mu=1$ component of the magnetic operator
$\hat F_{1\mu}^{\tau}$, which is much more
convenient for us. The terms $<{I_x^{\tau}}^2>$ are introduced to
ensure the orthogonality of the synthetic scissors to the spurious
state $|Sp>=(\hat I^{\rm n}+\hat I^{\rm p})|0>$.
However, we do not need these
terms because the collective states $|\nu>$ of our model are already
orthogonal to $|Sp>$ (see below); hence, the overlaps $<Syn|\nu>$
will be free from any admixtures of $|Sp>$. So, we use the
following definitions of the synthetic and spurious states:
$$|Syn>=\N^{-1}(\hat F_{11}^{\rm p}-\hat F_{11}^{\rm n})|0>,\qquad
|Sp>=(\hat F_{11}^{\rm p}+\hat F_{11}^{\rm n})|0>.$$

 Let us demonstrate the orthogonality
of the spurious state to all the rest of the states $|\nu>$.
As the first step it is necessary to show that the secular equation
(\ref{Secul1}) has the solution $E=0$.
We need the expression for $S^{\tau}(E=0)\equiv S^{\tau}(0)$.
In accordance with (\ref{S}), we have
$$S^{\tau}(E)=\left[\frac{\epsilon_0D_0}{E^2-\epsilon_0^2}+
\frac{\epsilon_2D_2}{E^2-\epsilon_2^2}\right]^{\tau},\quad
S^{\tau}(0)=-\left[\frac{D_0}{\epsilon_0}+
\frac{D_2}{\epsilon_2}\right]^{\tau}.$$
 The expressions for $D_0^{\tau},\,D_2^{\tau}$ are easily extracted
from formulae (B.10), (B.11):
\begin{equation}
D_0^{\tau}=\frac{\hbar}{m}Q_{00}^{\tau}\left[
\frac{1+\frac{4}{3}\delta}{\omega_x}
-\frac{1-\frac{2}{3}\delta}{\omega_z}\right]^{\tau},
\quad D_2^{\tau}=\frac{\hbar}{m}Q_{00}^{\tau}\left[
\frac{1+\frac{4}{3}\delta}{\omega_x}
+\frac{1-\frac{2}{3}\delta}{\omega_z}\right]^{\tau}.
\label{D0D2}
\end{equation}
So we find
\begin{eqnarray}
 S^{\tau}(0)&=&
-\frac{\hbar}{m}Q_{00}^{\tau}\left[
\frac{1+\frac{4}{3}\delta}{\omega_x}
(\frac{1}{\epsilon_2}+\frac{1}{\epsilon_0})
+\frac{1-\frac{2}{3}\delta}{\omega_z}
(\frac{1}{\epsilon_2}-\frac{1}{\epsilon_0})
\right]^{\tau}
\nonumber\\
&=&-\frac{\hbar^2}{m}
\frac{4\delta^{\tau}Q_{00}^{\tau}}
{\epsilon_2^{\tau}\epsilon_0^{\tau}}
=-\frac{1}{m}
\frac{3Q_{20}^{\tau}}
{(\omega_x^2-\omega_z^2)^{\tau}},
\label{S0}
\end{eqnarray}
where, in accordance with (B.12),
\begin{equation}
(\omega_x^2-\omega_z^2)^{\rm p}=
-\frac{6}{m}(\kappa Q_{20}^{\rm p}+\bar\kappa Q_{20}^{\rm n}),\quad
(\omega_x^2-\omega_z^2)^{\rm n}=
-\frac{6}{m}(\kappa Q_{20}^{\rm n}+\bar\kappa Q_{20}^{\rm p}).
\label{omeg-}
\end{equation}
Finally, we get
$$2S^{\rm p}(0)=\frac{Q_{20}^{\rm p}}{\kappa Q_{20}^{\rm p}
+\bar\kappa Q_{20}^{\rm n}},\quad
1-2S^{\rm p}(0)\kappa=\frac{\bar\kappa Q_{20}^{\rm n}}
{\kappa Q_{20}^{\rm p}+\bar\kappa Q_{20}^{\rm n}},$$
$$2S^{\rm n}(0)=\frac{Q_{20}^{\rm n}}{\kappa Q_{20}^{\rm n}
+\bar\kappa Q_{20}^{\rm p}},\quad
1-2S^{\rm n}(0)\kappa=\frac{\bar\kappa Q_{20}^{\rm p}}
{\kappa Q_{20}^{\rm n}+\bar\kappa Q_{20}^{\rm p}}.$$
It is easy to see that substituting these expressions into
(\ref{Secul1}) we obtain an identity; therefore, the secular
equation has a zero energy solution.

For the second step it is necessary to calculate the overlap
$<Sp|\nu>$.
 Summing (\ref{Magnet}) with an
analogous expression for neutrons, we get
\begin{eqnarray}
<Sp|\nu>
&=&\frac{\tilde e\hbar}{c\sqrt5}E_{\nu}\sum_{\tau}
K_{\nu}^{\tau}
(\omega_x^2-\omega_z^2)^{\tau}
\sum_{mi}\frac{|D^{\tau}_{mi}|^2}
{\epsilon_{mi}^{\tau}(E_{\nu}^2-\epsilon_{mi}^2)^{\tau}}
\nonumber\\
&=&\frac{\tilde e\hbar}{c\sqrt5}E_{\nu}\sum_{\tau}
K_{\nu}^{\tau}
(\omega_x^2-\omega_z^2)^{\tau}
\sum_{mi}\frac{|D^{\tau}_{mi}|^2\epsilon_{mi}^{\tau}}
{(\epsilon_{mi}^2)^{\tau}(E_{\nu}^2-\epsilon_{mi}^2)^{\tau}}.
\label{ortog}
\end{eqnarray}
Applying the algebraical identity
$$\frac{1}{\epsilon^2(E^2-\epsilon^2)}=
\frac{1}{E^2}(\frac{1}{\epsilon^2}+
\frac{1}{E^2-\epsilon^2})$$
and remembering the definition (\ref{S}) of $S^{\tau}$ we
rewrite (\ref{ortog}) as
\begin{eqnarray}
<Sp|\nu>
&=&\frac{\tilde e\hbar}{c\sqrt5E_{\nu}}\sum_{\tau}
K_{\nu}^{\tau}
(\omega_x^2-\omega_z^2)^{\tau}(S^{\tau}-S^{\tau}(0))
\nonumber\\
&=&\frac{\tilde e\hbar}{c\sqrt5}\frac{K_{\nu}^{\rm p}}{E_{\nu}}
\left[
(\omega_x^2-\omega_z^2)^{\rm p}
(S^{\rm p}-S^{\rm p}(0))
+(\omega_x^2-\omega_z^2)^{\rm n}
(S^{\rm n}-S^{\rm n}(0))
\frac{K_{\nu}^{\rm n}}{K_{\nu}^{\rm p}}
\right].
\label{ortogo}
\end{eqnarray}
In accordance with (\ref{CSK}) and (\ref{Cn/Cp}),
\begin{equation}
\frac{K_{\nu}^{\rm n}}{K_{\nu}^{\rm p}}=
\frac{1-2S^{\rm p}\kappa}{2S^{\rm n}\bar\kappa}.
\label{Kn/Kp}
\end{equation}
Noting now (see formula (\ref{S0})) that
$ (\omega_x^2-\omega_z^2)^{\tau}S^{\tau}(0)
=-\frac{3}{m}Q_{20}^{\tau}$ and
taking into account relations (\ref{omeg-}), we find
\begin{eqnarray}
<Sp|\nu>
&=&\beta
\left\{
[(\kappa Q_{2}^{\rm p}+\bar\kappa Q_{2}^{\rm n})2S^{\rm p}
-Q_{2}^{\rm p}]
+[(\kappa Q_{2}^{\rm n}+\bar\kappa Q_{2}^{\rm p})2S^{\rm n}
-Q_{2}^{\rm n}]
\frac{1-2S^{\rm p}\kappa}{2S^{\rm n}\bar\kappa}
\right\}
\nonumber\\
&=&\beta
\left\{
[(2S^{\rm p}\kappa-1)Q_{2}^{\rm p}
+2S^{\rm p}\bar\kappa Q_{2}^{\rm n}]
+[(2S^{\rm n}\kappa -1)Q_{2}^{\rm n}
+2S^{\rm n}\bar\kappa Q_{2}^{\rm p})]
\frac{1-2S^{\rm p}\kappa}{2S^{\rm n}\bar\kappa}
\right\}
\nonumber\\
&=&\beta
\left\{
2S^{\rm p}\bar\kappa Q_{2}^{\rm n}
+(2S^{\rm n}\kappa -1)Q_{2}^{\rm n}
\frac{1-2S^{\rm p}\kappa}
{2S^{\rm n}\bar\kappa}
\right\}
\nonumber\\
&=&\beta \frac{Q_{2}^{\rm n}}{2S^{\rm n}\bar\kappa}
\left\{
2S^{\rm n}\bar\kappa
2S^{\rm p}\bar\kappa
-(1-2S^{\rm n}\kappa)
(1-2S^{\rm p}\kappa)
\right\}=0,
\label{ortogon}
\end{eqnarray}
where $\di\beta=-\frac{3}{m}
\frac{\tilde e\hbar}{c\sqrt5}\frac{K_{\nu}^{\rm p}}{E_{\nu}}$
and $Q_2\equiv Q_{20}$.
The expression in the last curly brackets coincides obviously with the
secular equation (\ref{Secul1}) that proves the orthogonality of the
spurious state to all physical states of the considered model. So we
can conclude that strictly speaking this is not a spurious state,
but one of the exact eigenstates of the model corresponding to the
integral of motion $I^{\rm n}+I^{\rm p}$. In other words \cite{Ring}:
"In fact these excitations are not really spurious, but they represent
a different type of motion which has to be treated separately."
The same conclusion was made
by N. Lo Iudice \cite{Lo96} who solved this problem approximately
with the help of several assumptions (a small deformation limit, for
 example).

The problem of the "spurious" state being solved, the calculation of
the overlaps $<Syn|\nu>$ becomes trivial. Really, we have shown that
$<0|\hat F^{\rm n}_{11}+\hat F^{\rm p}_{11}|\nu>=0.$
That means that
$<0|\hat F^{\rm n}_{11}|\nu>=-<0|\hat F^{\rm p}_{11}|\nu>$; hence,
$<Syn|\nu>=\N^{-1}<0|\hat F^{\rm p}_{11}-\hat F^{\rm n}_{11}|\nu>=
2\N^{-1}<0|\hat F^{\rm p}_{11}|\nu>$ and
\begin{equation}
U^2\equiv|<Syn|\nu>|^2=2\N^{-2}B(M1)_{\nu}.
\label{U2}
\end{equation}
The nontrivial part of the problem is the calculation of the
normalization factor $\N$. It is important not to forget about the
time dependence of the synthetic state which should be determined
by the external field:
$$|Syn(t)>=\N^{-1}[
(\hat F_{11}^{\rm p}-\hat F_{11}^{\rm n})e^{-i\Omega t}
+(\hat F_{11}^{\rm p}-\hat F_{11}^{\rm n})^{\dagger}e^{i\Omega t}
]|0>.$$
 Then we have
\begin{eqnarray}
\N^2&=&2<0|(\hat F^{\rm p}_{11}-\hat F^{\rm n}_{11})^{\dagger}
(\hat F^{\rm p}_{11}-\hat F^{\rm n}_{11})|0>
\nonumber\\
&=&
2\sum_{ph}<0|(\hat F^{\rm p}_{11}-\hat F^{\rm n}_{11})^{\dagger}|ph>
<ph|(\hat F^{\rm p}_{11}-\hat F^{\rm n}_{11})|0>=
2\sum_{ph}|<ph|(\hat F^{\rm p}_{11}-\hat F^{\rm n}_{11})|0>|^2
\nonumber\\
&=&
2\sum_{\tau,ph}|<ph|\hat F^{\tau}_{11}|0>|^2=
2\sum_{\tau,ph}|(f^{\tau}_{11})_{ph}|^2.
\label{Norm}
\end{eqnarray}
With the help of relation (B.16) we find
\begin{eqnarray}
\N^2
&=&\frac{2}{5}
(\frac{e\hbar}{2c})^2 \sum_{\tau,ph} \left(
\omega_-^4
\frac{|<ph|r^2Y_{21}|0>|^2}{\epsilon^2_{ph}}\right)^{\tau}
\nonumber\\
&=&\frac{1}{8\pi}
(\frac{e\hbar}{2c})^2
 \sum_{\tau}
(\omega_-^4)^{\tau}
\left(
\frac{D_0}{\epsilon^2_0}
+\frac{D_2}{\epsilon^2_2}
\right)^{\tau}.
\label{Norma}
\end{eqnarray}
 Expressions for $D_0^{\tau},\,D_2^{\tau},\,
\omega_x^{\tau},\,\omega_z^{\tau}$ are given by formulae
(\ref{D0D2}), (B.12). To get a definite number, it is necessary
to make some assumption concerning the relation between neutron and
proton equilibrium characteristics. As usual, we apply the
approximation (\ref{Apr4}), i.e., suppose
$Q_{00}^{\rm n}=Q_{00}^{\rm p},\,Q_{20}^{\rm n}=Q_{20}^{\rm p}$.
It is easy to check that in this case formulae for
$\omega_{x,z}^{\tau}$ are reduced to the ones for the isoscalar case,
namely (A.3), and $D_0^{\tau}=D_0/2,\,D_2^{\tau}=D_2/2$, where
$D_0$ and $D_2$ are given by (\ref{sums}). So we get
\begin{eqnarray}
\N^2
=\frac{\omega_-^4}{8\pi}
(\frac{e\hbar}{2c})^2
\frac{Q_{00}}{m\bar\omega^2}
\left(\frac{1}{\epsilon_0}
+\frac{1}{\epsilon_2}\right)
=\frac{\delta}{2\pi}
\frac{m\omega_x}{\hbar}Q_{00}\mu_N^2.
\label{Norma2}
\end{eqnarray}
The estimation of the overlap for $^{156}$Gd with $\delta=0.27$ gives
$\N^2=34.72\mu_N^2$ and $U^2=0.53$ (see eq. (\ref{U2})), that is two
times larger than the
result of \cite{Hilt86} obtained in QRPA calculations with the Skyrme
forces. The disagreement can naturally be attributed to the difference
in forces and especially to the lack of pair correlations in our
approach. In a small deformation
 limit $U^2=\frac{1}{2}\sqrt{\frac{3}{2}}\approx 0.6.$

 This is the maximum possible overlap of the "pure" (or "synthetic")
scissors with the real scissors. The increasing of  $\delta$ and /or
taking into account pairing correlations decreases its value, that is
confirmed by numerous microscopic calculations with various forces
\cite{Zaw}. Such small overlap leads inevitably to the conclusion,
that the original model of counter rotating rigid rotors \cite{Lo2000}
has not very much in common with the real scissors mode, the correct
description of which requires the proper treatment of the Fermi surface
deformation and the coupling with IVGQR.

\subsubsection{Superdeformation}

A certain drawback of our approach is that, so
far, we have not included the superfluidity into our description.
Nevertheless, our formulae (\ref{Omeg2fin}, \ref{BM1gen}) can be
successfully used
for the description of superdeformed nuclei where the pairing
is very weak \cite{Hamam,Lo2000}. For example, applying them to the
superdeformed nucleus $^{152}$Dy ($\delta\simeq 0.6,\,
\hbar\omega_0=41/A^{1/3} \rm{MeV}$), we get
$$E_{iv}=20.8\, {\rm MeV},\quad\quad B(M1)_{iv}=15.9\, \mu_N^2$$
for the isovector GQR and
$$E_{sc}=4.7\, {\rm MeV},\quad\quad B(M1)_{sc}=20.0\, \mu_N^2$$
for the scissors mode. There are not so many results of other
calculations to compare with. As a matter of fact, there are only two
papers considering this problem.

The phenomenological TRM model \cite{Lo2000} predicts
$$E_{iv}\simeq26\, {\rm MeV},\quad B(M1)_{iv}\simeq26\, \mu_N^2,\quad
E_{sc}\simeq6.1\, {\rm MeV},\quad B(M1)_{sc}\simeq22\, \mu_N^2.$$
The only existing microscopic calculation \cite{Hamam}
in the framework of QRPA with separable forces gives
$$E_{iv}\simeq28\, {\rm MeV},\quad B(M1)_{iv}\simeq37\, \mu_N^2,\quad
E_{sc}\simeq5-6\, {\rm MeV},\quad B(M1)_{1^+}\simeq23\, \mu_N^2.$$
Here $B(M1)_{1^+}$ denotes
the total $M1$ orbital strength carried by the
calculated $K^{\pi}=1^+$ QRPA excitations modes in the energy region
below 20 MeV.

It is easy to see that in the case of IVGQR one can speak, at least,
about qualitative agreement. Our
results for $E_{sc}$ and $B(M1)_{sc}$ are in good agreement with
that of phenomenological model and with $E_{sc}$ and $B(M1)_{1^+}$ of
Hamamoto and Nazarewicz.

 It is possible to extract from the histogram
of \cite{Hamam} the value of the overlap of calculated low-lying
$1^+$ excitations with the synthetic scissors state:
$|<Syn|1^+>|^2 \approx 0.4.$ The result of our calculation
$U^2=0.43$ agrees with it very well. So the natural conclusion of
this section is that the correct treatment of pair correlations is
obligatory for a reasonable description of the scissors mode.

\subsection{Equations of motion}

Let us look on WFM equations of motion from the RPA point of view.
Is it possible to construct something similar in the RPA approach?
Equations (\ref{quadr}) are written for average values of
operators and are valid for the description of the arbitrary
amplitude motion. One can compare with RPA only their linearised
version, obtained by the variation of equations. The variables of
linearised equations are the variations of the above mentioned
average values. It is natural to suppose some correspondence
between the variation of the average value of $\hat F$ operator and
the matrix element of the type $<0|\hat F|\nu>$ used to
calculate transition probabilities. To check this idea we have to
derive dynamical equations for matrix elements of the operators
$r^2_{\lambda\mu}$, $\hat p^2_{\lambda\mu}$ and
$(r\hat p)_{\lambda\mu}$ to compare them with linearised equations
(\ref{quadr}). To this end we combine RPA equations (\ref{DD}) in
accordance with the definition (\ref{matelem}) of matrix elements:
\begin{equation}
\label{Equmot}
\hbar\Omega_{\nu}\sum_{mi}(f_{im}^{\tau}X_{mi}^{\tau,{\nu}}
+f_{mi}^{\tau}Y_{mi}^{\tau,{\nu}})=
\sum_{mi}\epsilon_{mi}(f_{im}^{\tau}X_{mi}^{\tau,{\nu}}
-f_{mi}^{\tau}Y_{mi}^{\tau,{\nu}})+
K^{\tau}_{\nu}\sum_{mi}(f_{im}^{\tau}D_{im}^{\tau*}
-f_{mi}^{\tau}D_{mi}^{\tau*}).
\end{equation}
 Taking into account the relations
$$\epsilon_{mi}f_{im}=[\hat f,H_0]_{im},\quad
\epsilon_{mi}f_{mi}=-[\hat f,H_0]_{mi},\quad$$
one rewrites this equation as
\begin{equation}
\label{Equmot2}
\hbar\Omega_{\nu}<0|\hat F^{\tau}|\nu>=
\sum_{mi}\{[\hat f^{\tau},H_0^{\tau}]_{im}X_{mi}^{\tau,{\nu}}
+[\hat f^{\tau},H_0^{\tau}]_{mi}Y_{mi}^{\tau,{\nu}}+
K^{\tau}_{\nu}(f_{im}^{\tau}D_{im}^{\tau*}
-f_{mi}^{\tau}D_{mi}^{\tau*})\}.
\end{equation}
The Hamiltonian of the axially deformed harmonic oscillator
corresponding to the mean field (\ref{potenirr}) is
\begin{equation}
\label{H0}
H_0^{\tau}(\br)=\sum_{s=1}^{N_{\tau}}\{\frac{\hat\bp_s^2}{2m}+
\frac{1}{2}m\,\omega^2\br_s^2+
Z_{20}^{\tau}(eq)r^2_{20}(s)\}.
\end{equation}

Let us consider the operator
$\hat f=\sqrt6\,r^2_{21}=q_{21}=D.$
Calculating the commutator
$$[r^2_{21},H_0]=i\hbar\frac{2}{m}(r\hat p)_{21}$$
we find from (\ref{Equmot2}) the following equation
\begin{eqnarray}
\hbar\Omega_{\nu}<0|\sum_{s=1}^{N_{\tau}}D_s^{\tau}|\nu>&=&
i\hbar\sqrt6\frac{2}{m}
\sum_{mi}
\{((r\hat p)_{21})_{im}^{\tau}X_{mi}^{\tau,{\nu}}+
((r\hat p)_{21})_{mi}^{\tau}Y_{mi}^{\tau,{\nu}}\}
\nonumber\\
&&+K^{\tau}_{\nu}\sum_{mi}
(D_{im}^{\tau}D_{im}^{\tau*}
-D_{mi}^{\tau}D_{mi}^{\tau*}).
\label{EquD}
\end{eqnarray}
Taking into account relations $(D^*)_{im}=(D)_{mi}^*$
and $|D_{mi}|^2=|D_{im}|^2$ we find, that the last sum in (\ref{EquD})
is equal to zero. Applying again formula (\ref{matelem}) and
introducing the notation
$\di\hat R_{\lambda\mu}=\sum_{s=1}^A(r^2_s)_{\lambda\mu},\quad
\hat L_{\lambda\mu}=\sum_{s=1}^A(r_s\hat p_s)_{\lambda\mu}$
we write (\ref{EquD}) as
\begin{equation}
-i\Omega_{\nu}<0|\hat R_{21}^{\tau}|\nu>
=\frac{2}{m}<0|\hat L_{21}^{\tau}|\nu>.
\label{Dyneq}
\end{equation}
 Identifying the matrix elements
$<0|\hat R_{21}^{\tau}|\nu>$ and
$<0|\hat L_{21}^{\tau}|\nu>$
with $\R^{\tau}_{21}$ and $\L^{\tau}_{21}$ respectively we
reproduce the variation of the first equation in (\ref{quadr})
(having in mind the time dependence via $e^{-i\Omega t}$).

 Let us consider the operator $\hat f=(r\hat p)_{21}.$
The required commutator is evaluated to be
$$[(r\hat p)_{21},H_0]=i\frac{\hbar}{m}\hat p^2_{21}
-i\hbar m\omega^2r^2_{21}-i\frac{\hbar}{\sqrt6}
Z_{20}(eq)r^2_{21}.$$
With this result equation (\ref{Equmot2}) looks as
\begin{eqnarray}
\label{rp21equ}
\hbar\Omega_{\nu}<0|\hat L_{21}^{\tau}|\nu>=
i\frac{\hbar}{m}<0|\hat P_{21}^{\tau}|\nu>
-i\hbar m\omega^2<0|\hat R_{21}^{\tau}|\nu>
\nonumber\\
-i\frac{\hbar}{\sqrt6}
Z_{20}^{\tau}(eq)<0|\hat R_{21}^{\tau}|\nu>+
K^{\tau}_{\nu}\sum_{mi}[((r\hat p)_{21}^{\tau})_{im}D_{im}^{\tau*}
-((r\hat p)_{21}^{\tau})_{mi}D_{mi}^{\tau*}],
\end{eqnarray}
where the notation
$\di\hat P_{\lambda\mu}=\sum_{s=1}^A(\hat p^2_s)_{\lambda\mu}$
has been introduced.
The last sum is calculated with the help of formula (B.17).
Using the fact, that $\epsilon_{im}=-\epsilon_{mi}$, one gets
\begin{eqnarray}
&&\sum_{mi}[((r\hat p)_{21})_{im}^{\tau}D_{im}^{\tau*}
-((r\hat p)_{21})_{mi}^{\tau}D_{mi}^{\tau*}]=
-i\frac{m}{2\hbar}
\sum_{mi}\epsilon_{mi}[(r^2_{21})_{im}^{\tau}D_{im}^{\tau*}
+(r^2_{21})_{mi}^{\tau}D_{mi}^{\tau*}]
\nonumber\\
&&=-i\frac{m}{\hbar\sqrt6}
\sum_{mi}\epsilon_{mi}^{\tau}|D_{mi}^{\tau}|^2=
-i\frac{m}{\hbar\sqrt6}
(\epsilon_{0}^{\tau}D_{0}^{\tau}
+\epsilon_{2}^{\tau}D_{2}^{\tau})=
-i\hbar\frac{4}{\sqrt6}(1+\delta/3)Q_{00}^{\tau}.
\nonumber
\end{eqnarray}
According to the definitions (see formula \ref{CSK}) we have
$$K^{\rm n}_{\nu}=\sum_{\tau}\kappa_{\rm n\tau}C^{\tau}_{\nu}
=(\chi<0|\hat R_{21}^{\rm n}|\nu>
+\bar \chi<0|\hat R_{21}^{\rm p}|\nu>)/\sqrt6,$$
$$K^{\rm p}_{\nu}=\sum_{\tau}\kappa_{\rm p\tau}C^{\tau}_{\nu}
=(\chi<0|\hat R_{21}^{\rm p}|\nu>
+\bar \chi<0|\hat R_{21}^{\rm n}|\nu>)/\sqrt6.$$
So, the equation (\ref{rp21equ}) (let us say, for neutrons) is
transformed into
\begin{eqnarray}
\label{rp21eq}
&&-i\Omega_{\nu}<0|\hat L_{21}^{\rm n}|\nu>=
+\frac{1}{m}<0|\hat P_{21}^{\rm n}|\nu>
-m\omega^2<0|\hat R_{21}^{\rm n}|\nu>
\nonumber\\
&&-\frac{1}{\sqrt6}Z_{20}^{\rm n}(eq)<0|\hat R_{21}^{\rm n}|\nu>
-\frac{2}{3}(1+\delta/3)Q_{00}^{\rm n}
(\chi<0|\hat R_{21}^{\rm n}|\nu>
+\bar \chi<0|\hat R_{21}^{\rm p}|\nu>).
\end{eqnarray}
The equation for protons is obtained by interchanging indices n and
p. One has to compare this equation with the variation of the second
equation in (\ref{quadr}) with $\lambda=2,\, \mu=1$. Let us write
this variation in detail:
$$\frac{d}{dt}\L^{\tau}_{21}
-\frac{1}{m}\P_{21}^{\tau}+
m\,\omega^2\R^{\tau}_{21}
-2\sqrt{5}\sum_{j=0,2}\sqrt{2j+1}
\{_{221}^{11j}\}
\sum_{\sigma,\nu}C_{2\sigma,j\nu}^{21}
[Z_{2\sigma}^{\tau}(eq)\R_{j\nu}^{\tau}+
\delta Z_{2\sigma}^{\tau}R_{j\nu}^{\tau}(eq)]=0.$$
We recall, that only $R_{00}^{\tau}(eq)$ and $R_{20}^{\tau}(eq)$ have
non zero values, so this equation is reduced to
$$\frac{d}{dt}\L^{\tau}_{21}
-\frac{1}{m}\P_{21}^{\tau}+
m\,\omega^2\R^{\tau}_{21}
-10\{_{221}^{112}\}C_{20,21}^{21}
Z_{20}^{\tau}(eq)\R_{21}^{\tau}$$
$$-2\sqrt5\delta Z_{21}^{\tau}[
\{_{221}^{110}\}C_{21,00}^{21}R_{00}^{\tau}(eq)
+\sqrt5\{_{221}^{112}\}C_{21,20}^{21}R_{20}^{\tau}(eq)]=0.$$
In agreement with definition (\ref{potenirr}) of
$Z_{\lambda\mu}^{\tau}$ its variation is
$$ \delta Z_{2\mu}^{\rm n}=\chi \R_{2\mu}^{\rm n}
+\bar{\chi}\R_{2\mu}^{\rm p}\,,\quad
\delta Z_{2\mu}^{\rm p}=\chi \R_{2\mu}^{\rm p}
+\bar{\chi}\R_{2\mu}^{\rm n}\,.$$
Substituting 6j-symbols and Clebsch-Gordan coefficients by their
numerical values we obtain finally (e.g. for neutrons)
$$\frac{d}{dt}\L^{\rm n}_{21}
-\frac{1}{m}\P_{21}^{\rm n}+
m\,\omega^2\R^{\rm n}_{21}
+\frac{1}{\sqrt6}Z_{20}^{\rm n}(eq)\R_{21}^{\rm n}
+\frac{2}{3}(1+\delta/3)Q_{00}^{\rm n}
(\chi\R_{21}^{\rm n}
+\bar\chi\R_{21}^{\rm p})=0.$$
This equation coincides obviously with (\ref{rp21eq}) if to
assume the time dependence via $e^{-i\Omega t}$ and to identify
the matrix elements $<0|\hat R_{21}^{\tau}|\nu>$,
$<0|\hat L_{21}^{\tau}|\nu>$ and
$<0|\hat P_{21}^{\tau}|\nu>$
with the variables $\R^{\tau}_{21}$,
$\L^{\tau}_{21}$ and $\P^{\tau}_{21}$,
respectively.

Let us consider the operator $\hat f=(\hat p^2)_{21}.$
The required commutator is
$$[(\hat p^2)_{21},H_0]=-i\hbar 2m\omega^2(r\hat p)_{21}
+i\hbar 4\sqrt5\sum_{j=1}^2\sqrt{2j+1}\{_{221}^{11j}\}C_{20,j1}^{21}
Z_{20}(eq)(r\hat p)_{j1}$$
and one obtains from (\ref{Equmot2}) the following equation
\begin{eqnarray}
\hbar\Omega_{\nu}<0|\hat P_{21}^{\tau}|\nu>&=&
-i\hbar 2m\omega^2<0|\hat L_{21}^{\tau}|\nu>
\nonumber\\
&&+i\hbar 4\sqrt5\sum_{j=1}^2\sqrt{2j+1}\{_{221}^{11j}\}C_{20,j1}^{21}
Z_{20}^{\tau}(eq)<0|\hat L_{j1}^{\tau}|\nu>
\nonumber\\
&&+K^{\tau}_{\nu}\sum_{mi}[((\hat p^2)_{21}^{\tau})_{im}D_{im}^{\tau*}
-((\hat p^2)_{21}^{\tau})_{mi}D_{mi}^{\tau*}].
\label{pp21equ}
\end{eqnarray}
It is easy to show (with the help of formula (B.18) that the
last sum is equal to zero.
This equation must be compared with the variation of the last
equation in (\ref{quadr}) with $\lambda=2,\, \mu=1$. Let us write it
in detail. Taking into account that $L^{\tau}_{\lambda \mu}(eq)=0$
we find the equation
\begin{eqnarray}
\frac{d}{dt}\P_{21}^{\tau}
+2m\,\omega^2\L^{\tau}_{21}
-4\sqrt5\sum_{j=1}^2\sqrt{2j+1}\{_{221}^{11j}\}C_{20,j1}^{21}
Z_{20}^{\tau}(eq)\L^{\tau}_{j1}=0
\nonumber
\end{eqnarray}
that obviously coincides with (\ref{pp21equ}) if to assume the
$e^{-i\Omega t}$ time dependence and to identify the
proper RPA matrix elements with the respective WFM variables.

\section{WFM versus RPA}

The exact relation between RPA matrix elements and the respective WFM
variables can be established with the help of the linear responce
theory. Let us first recall, following Appendix D of
\cite{Ring}, the necessary definitions concerning the density
and the density matrix.

{\bf The density operator} is defined as

\begin{equation}
\label{densop}
\hat\rho(\br)=\sum_{s=1}^A\delta(\br-\hat\br_s)=
\sum_{pq}d_{pq}(\br)a_p^{\dagger}a_q,
\end{equation}
where $\di{d_{pq}(\br)=<p|\delta(\br-\hat\br)|q>
=\sum_{\sigma,\tau}\phi_p^*(\br\sigma\tau)\phi_q(\br\sigma\tau)}$
and $\phi_q(\br\sigma\tau)$ are single-particle
wave functions. Indices $p,q$ include spin and isospin quantum
numbers $\sigma$ and $\tau$.

{\bf The density of particles} in the system depends on its state
$\Psi$ and is defined as the average value of a density
operator over this state:
\begin{eqnarray}
&&\rho(\br)=<\Psi|\hat\rho(\br)|\Psi>
=\sum_{pq}d_{pq}(\br)\rho_{qp}
\nonumber\\
&&=A\sum_{\sigma,\tau,...,\sigma_A,\tau_A}
\int d^3r_2...d^3r_A|\Psi(\br\sigma\tau,\br_2\sigma_2\tau_2,
...,\br_A\sigma_A\tau_A)|^2,
\label{density}
\end{eqnarray}
where $\rho_{qp}=<\Psi|a_p^{\dagger}a_q|\Psi>.$
The particle density (\ref{density}) can be interpreted as the diagonal
element (in the coordinate space representation) of {\bf the density
matrix} which is defined as
\begin{equation}
\label{densmat}
\rho(\br\sigma\tau,\br'\sigma'\tau')
=\sum_{pq}\phi_p^*(\br'\sigma'\tau')\phi_q(\br\sigma\tau)
<\Psi|a_p^{\dagger}a_q|\Psi>
=\sum_{pq}d_{pq}(\br'\sigma'\tau',\br\sigma\tau)\rho_{qp}
\end{equation}
with $d_{pq}(\br'\sigma'\tau',\br\sigma\tau)
=\phi_p^*(\br'\sigma'\tau')\phi_q(\br\sigma\tau)$.
 The average value of the arbitrary one-body operator
\begin{equation}
\label{Foper}
 \hat F=\sum_{s=1}^A\hat f_s=\sum_{pq}f_{pq}a^{\dagger}_p a_q
\end{equation}
is written in terms of the density matrix as
$$<\Psi|\hat F|\Psi>=\sum_{pq}f_{pq}<\Psi|a_p^{\dagger}a_q|\Psi>
=\sum_{pq}f_{pq}\rho_{qp}=Tr(f\rho).$$

 Let us consider the system to be
in the weak external time-dependent field
\begin{equation}
\label{Exfield}
\hat W(t)=\hat W\,{\rm exp}(-i\Omega t)
+\hat W^{\dagger}\,{\rm exp}(i\Omega t),
\end{equation}
where $\hat W=\sum_{pq}w_{pq}a^{\dagger}_p a_q$ is a one-body operator.
The change of the ground state wave function produced by this field is
found by using the time-dependent perturbation theory \cite{LL}:
\begin{eqnarray}
\Psi(t)=
|0>+\sum_{\nu}|\nu>\left[c_{\nu}e^{-i\Omega t}
-\bar c_{\nu}^*e^{i\Omega t}\right].
\label{Psi}
\end{eqnarray}
Here $|0>$ and $|\nu>$ are stationary eigenstates of the unperturbed
system and
\begin{equation}
\label{Cnu}
 c_{\nu}=\frac{<\nu|\hat W|0>}{\hbar(\Omega-\Omega_{\nu})}
=\sum_{pq}
\frac{<\nu|a^{\dagger}_p a_q|0>}{\hbar(\Omega-\Omega_{\nu})}
w_{pq},\quad
\bar c_{\nu}=\frac{<0|\hat W|\nu>}{\hbar(\Omega+\Omega_{\nu})}
=\sum_{pq}
\frac{<0|a^{\dagger}_p a_q|\nu>}{\hbar(\Omega+\Omega_{\nu})}
w_{pq}.
\end{equation}
 Inserting this expression into formula
(\ref{densmat}) we obtain the perturbed density matrix
$$\rho(\br\sigma\tau,\br'\sigma'\tau',t)=
\rho_0(\br\sigma\tau,\br'\sigma'\tau')
+\delta\rho(\br\sigma\tau,\br'\sigma'\tau',t),$$
where $\rho_0(\br\sigma\tau,\br'\sigma'\tau')$ is the unperturbed
(equilibrium) density matrix
$$\rho_0(\br\sigma\tau,\br'\sigma'\tau')
=\sum_{pq}d_{qp}(\br'\sigma'\tau',\br\sigma\tau)<0|a_q^{\dagger}a_p|0>
=\sum_{pq}d_{qp}(\br'\sigma'\tau',\br\sigma\tau)\rho^{(0)}_{pq}$$
and $\delta\rho(\br\sigma\tau,\br'\sigma'\tau',t)$ is the change of
the density matrix
\begin{equation}
\delta\rho(\br\sigma\tau,\br'\sigma'\tau',t)
=\sum_{pq}d_{qp}(\br'\sigma'\tau',\br\sigma\tau)\rho^{(1)}_{pq}(t)
\label{chandens}
\end{equation}
with
\begin{eqnarray}
\rho^{(1)}_{pq}(t)&=&
\sum_{\nu}\left[
(<0|a_q^{\dagger}a_p|\nu>c_{\nu}
-<\nu|a_q^{\dagger}a_p|0>\bar c_{\nu})e^{-i\Omega t}
\right.
\nonumber\\
&&\left.
+(<\nu|a_q^{\dagger}a_p|0>c_{\nu}^*
-<0|a_q^{\dagger}a_p|\nu>\bar c_{\nu}^*)
e^{i\Omega t}\right].
\label{deltarho1}
\end{eqnarray}
Deriving (\ref{chandens}) we neglected the terms proportional to
$|\hat W|^2$.
At this stage it is necessary to remind that we work in a Hartree-Fock
approximation. That means that stationary states $|0>, |\nu>$ are
Slater determinants; matrix $\rho^{(0)}_{pq}=\rho_q\delta_{pq}$ is
diagonal with $\rho_q=1$ for levels below the Fermi level and
$\rho_q=0$ for levels above the Fermi level. The requirement
$(\rho_0+\delta\rho)^2=(\rho_0+\delta\rho)$ leads to the well known
\cite{Ring} property of the matrix $\rho^{(1)}_{pq}$: it has only
particle--hole nonvanishing matrix elements. Looking to formula
(\ref{deltarho1}) we see that it is possible for the matrix elements
$<0|a_q^{\dagger}a_p|\nu>$ to be different from zero only for
particle--hole combinations of indices $q,p$. Consequently, the
summation over $p,q$ in formula (\ref{Cnu}) for $c_n$ and $\bar c_n$
will also be restricted only to particle--hole pairs. So we
can write $\rho^{(1)}_{pq}$ as
$$\rho^{(1)}_{pq}(t)=
\sum_{p'q'}\left[R_{pq,p'q'}(\Omega)e^{-i\Omega t}
+R_{qp,p'q'}^*(\Omega)e^{i\Omega t}\right]w_{p'q'},$$
where
$$R_{pq,p'q'}(\Omega)=\sum_{\nu}\left(
\frac{<0|a^{\dagger}_q a_p|\nu><\nu|a^{\dagger}_{p'} a_{q'}|0>}
{\hbar(\Omega-\Omega_{\nu})}-
\frac{<0|a^{\dagger}_{p'} a_{q'}|\nu><\nu|a^{\dagger}_{q} a_{p}|0>}
{\hbar(\Omega+\Omega_{\nu})}\right)$$
is the RPA response function \cite{Ring}, where the index pairs $pq$
and $p'q'$ are restricted to particle--hole pairs. For the change
of the arbitrary operator average value we have:
\begin{equation}
\label{varF}
\delta<\Psi|\hat F|\Psi>=\sum_{pq}f_{pq}\rho_{qp}^{(1)}.
\end{equation}

Now we are ready to analyze the WFM variables. The first one is
$$ R_{\lambda\mu}^{\tau}(t)=
2(2\pi\hbar)^{-3}\int\! d^3p\,\int\! d^3r
r_{\lambda\mu}^{2}f^{\tau}(\br,\bp,t).$$
Using here the definition (\ref{f}) of the Wigner function and the
definition of the $\delta$-function we find
\begin{eqnarray}
 R_{\lambda\mu}^{\tau}(t)&=&
\frac{2}{(2\pi\hbar)^{3}}
\int\! d^3r\,
r_{\lambda\mu}^{2} \int d^3s\int\! d^3p
\: \exp(- i\bp\cdot
 \bs/\hbar)\rho^{\tau}(\br+\frac{\bs}{2}, \br-\frac{\bs}{2},t)
\nonumber\\
&=&2 \int\! d^3r\, r_{\lambda\mu}^{2}\rho^{\tau}(\br, \br,t)
=\sum_{\sigma} \int\! d^3r\, r_{\lambda\mu}^{2}
\rho(\br\sigma\tau,\br\sigma\tau,t)
\nonumber\\
&=&\sum_{pq}\sum_{\sigma} \int\! d^3r\, r_{\lambda\mu}^{2}
\phi_p^*(\br\sigma\tau)\phi_q(\br\sigma\tau)
<\Psi|a_p^{\dagger}a_q|\Psi>
\nonumber\\
&=&\sum_{pq}(r_{\lambda\mu}^{2})_{pq}^{\tau}\rho_{qp}(t)
=<\Psi|\sum_{pq}(r_{\lambda\mu}^{2})_{pq}^{\tau}a_p^{\dagger}a_q|\Psi>
\nonumber\\
&=&<\Psi|\sum_{s=1}^{N_{\tau}} r_{\lambda\mu}^{2}(s)|\Psi>
=<\Psi|\hat R_{\lambda\mu}^{\tau}|\Psi>.
\label{R2rho}
\end{eqnarray}
For the variation of this variable one can write the following chain
of relations
\begin{eqnarray}
 \delta R_{\lambda\mu}^{\tau}(t)&=&
 \R_{\lambda\mu}^{\tau}(t)=
2\int\! d^3r\, r_{\lambda\mu}^{2}\delta\rho^{\tau}(\br, \br,t)
\nonumber\\
&=&\sum_{pq}\sum_{\sigma} \int\! d^3r\, r_{\lambda\mu}^{2}
\phi_p^*(\br\sigma\tau)\phi_q(\br\sigma\tau)\rho^{(1)}_{qp}(t)
=\sum_{pq}(r_{\lambda\mu}^{2})_{pq}^{\tau}\rho^{(1)}_{qp}(t)
\nonumber\\
&=&\sum_{\nu}(<0|\hat R_{\lambda\mu}^{\tau}|\nu>c_{\nu}
-<\nu|\hat R_{\lambda\mu}^{\tau}|0>\bar c_{\nu})e^{-i\Omega t}
\nonumber\\
&&+\sum_{\nu}(<\nu|\hat R_{\lambda\mu}^{\tau}|0>c_{\nu}^*
-<0|\hat R_{\lambda\mu}^{\tau}|\nu>\bar c_{\nu}^*)e^{i\Omega t}.
\label{deltaR}
\end{eqnarray}

For the second variable we have
\begin{eqnarray}
 L_{\lambda\mu}^{\tau}(t)&=&
2(2\pi\hbar)^{-3}\int\! d^3p\,\int\! d^3r
(rp)_{\lambda\mu}f^{\tau}(\br,\bp,t)
\nonumber\\
&=&\frac{2}{(2\pi\hbar)^{3}}
\int\! d^3r\,\int\! d^3s\int\! d^3p
\: (rp)_{\lambda\mu} \exp(- i\bp\cdot
 \bs/\hbar)\rho^{\tau}(\br+\frac{\bs}{2}, \br-\frac{\bs}{2},t)
\nonumber\\
&=&-i\hbar \int\! d^3r\, \{(r[\nabla-\nabla'])_{\lambda\mu}
\rho^{\tau}(\br, \br',t)\}_{r=r'}
\nonumber\\
&=&-\frac{i\hbar}{2}\sum_{\sigma} \int\! d^3r\,
\{(r[\nabla-\nabla'])_{\lambda\mu}
\rho(\br\sigma\tau, \br'\sigma\tau,t)\}_{r=r'}
\nonumber\\
&=&-\frac{i\hbar}{2} \sum_{pq}\sum_{\sigma} \int\! d^3r\,
\{\phi_p^*(\br\sigma\tau)(r\nabla)_{\lambda\mu}\phi_q(\br\sigma\tau)
\nonumber\\
&&-\phi_q(\br\sigma\tau)(r\nabla)_{\lambda\mu}\phi_p^*(\br\sigma\tau)\}
<\Psi|a_p^{\dagger}a_q|\Psi>
\nonumber\\
&=&\sum_{pq}
\{((r\hat p)_{\lambda\mu})_{pq}^{\tau}
+i\hbar\frac{\sqrt3}{2}\delta_{\lambda 0}\sum_{\sigma}\int\! d^3r\,
\phi_p^*(\br\sigma\tau)\phi_q(\br\sigma\tau)\}\rho_{qp}(t)
\nonumber\\
&=&<\Psi|\sum_{s=1}^{N_{\tau}} (r_s\hat p_s)_{\lambda\mu}|\Psi>
+i\hbar\frac{\sqrt3}{2}\delta_{\lambda 0}N_{\tau}
=<\Psi|\hat L_{\lambda\mu}^{\tau}|\Psi>
+i\hbar\frac{\sqrt3}{2}\delta_{\lambda 0}N_{\tau}.
\label{Lrho}
\end{eqnarray}

The variation of this variable is
\begin{eqnarray}
 \delta L_{\lambda\mu}^{\tau}(t)
&=&\L_{\lambda\mu}^{\tau}(t)
=-i\hbar \int\! d^3r\, \{(r[\nabla-\nabla'])_{\lambda\mu}
\delta\rho^{\tau}(\br, \br',t)\}_{r=r'}
\nonumber\\
&=&\sum_{pq}((r\hat p)_{\lambda\mu})_{pq}^{\tau}\rho^{(1)}_{qp}(t)
\nonumber\\
&=&\sum_{\nu}[<0|\hat L_{\lambda\mu}^{\tau}|\nu>c_{\nu}
-<\nu|\hat L_{\lambda\mu}^{\tau}|0>\bar c_{\nu}]e^{-i\Omega t}
\nonumber\\
&&+\sum_{\nu}[<\nu|\hat L_{\lambda\mu}^{\tau}|0>c_{\nu}^*
-<0|\hat L_{\lambda\mu}^{\tau}|\nu>\bar c_{\nu}^*]e^{i\Omega t}.
\label{deltaL}
\end{eqnarray}

The third variable is
\begin{eqnarray}
 P_{\lambda\mu}^{\tau}(t)&=&
2(2\pi\hbar)^{-3}\int\! d^3p\,\int\! d^3r
p_{\lambda\mu}^{2}f^{\tau}(\br,\bp,t)
\nonumber\\
&=&\frac{2}{(2\pi\hbar)^{3}}
\int\! d^3r \int\! d^3s \int\! d^3p\, p^2_{\lambda\mu}
\: \exp(- i\bp\cdot
 \bs/\hbar)\rho^{\tau}(\br+\frac{\bs}{2}, \br-\frac{\bs}{2},t)
\nonumber\\
&=&-\frac{\hbar^2}{2} \int\! d^3r\, \{(\nabla-\nabla')^2_{\lambda\mu}
\rho^{\tau}(\br, \br',t)\}_{r=r'}
\nonumber\\
&=&-\frac{\hbar^2}{4}\sum_{\sigma} \int\! d^3r\,
\{(\nabla-\nabla')^2_{\lambda\mu}
\rho(\br\sigma\tau, \br'\sigma\tau,t)\}_{r=r'}
\nonumber\\
&=&-\frac{\hbar^2}{4} \sum_{pq}\sum_{\sigma} \int\! d^3r\,
\{\phi_p^*(\br\sigma\tau)\nabla^2_{\lambda\mu}\phi_q(\br\sigma\tau)
+\phi_q(\br\sigma\tau)\nabla^2_{\lambda\mu}\phi_p^*(\br\sigma\tau)
\nonumber\\
&&-2[\nabla\phi_q(\br\sigma\tau)\nabla
\phi_p^*(\br\sigma\tau)]_{\lambda\mu}\}
<\Psi|a_p^{\dagger}a_q|\Psi>
\nonumber\\
&=&-\hbar^2 \sum_{pq}\sum_{\sigma} \int\! d^3r\,
\phi_p^*(\br\sigma\tau)\nabla^2_{\lambda\mu}\phi_q(\br\sigma\tau)
<\Psi|a_p^{\dagger}a_q|\Psi>
\nonumber\\
&=&\sum_{pq}(\hat p^2_{\lambda\mu})_{pq}^{\tau}\rho_{qp}(t)
=<\Psi|\sum_{s=1}^{N_{\tau}} \hat p^2_{\lambda\mu}(s)|\Psi>
=<\Psi|\hat P_{\lambda\mu}^{\tau}|\Psi>.
\label{Prho}
\end{eqnarray}

The variation of this variable is
\begin{eqnarray}
 \delta P_{\lambda\mu}^{\tau}(t)
&=& \P_{\lambda\mu}^{\tau}(t)=
-\frac{\hbar^2}{2}\int\! d^3r\, \{(\nabla-\nabla')^2_{\lambda\mu}
\delta\rho^{\tau}(\br, \br',t)\}_{r=r'}
\nonumber\\
&=&\sum_{pq}
(\hat p^2_{\lambda\mu})_{pq}^{\tau}\rho^{(1)}_{qp}(t)
\nonumber\\
&=&\sum_{\nu}[<0|\hat P_{\lambda\mu}^{\tau}|\nu>c_{\nu}
-<\nu|\hat P_{\lambda\mu}^{\tau}|0>
\bar c_{\nu}]e^{-i\Omega t}
\nonumber\\
&&+\sum_{\nu}[<\nu|\hat P_{\lambda\mu}^{\tau}|0>c_{\nu}^*
-<0|\hat P_{\lambda\mu}^{\tau}|\nu>\bar c_{\nu}^*]e^{i\Omega t}.
\label{deltaP}
\end{eqnarray}

The structure of variables
$\R_{\lambda\mu}$, $\L_{\lambda\mu}$, $\P_{\lambda\mu}$
(\ref{deltaR},\ref{deltaL},\ref{deltaP}) demonstrates in an obvious
way the relation between the WFM method and RPA. One sees, for
example, that the dynamical equations for the WFM variables
$\R_{\lambda\mu}$ is a linear combination of the dynamical equations
(\ref{Dyneq}) for the transition matrix elements
$<0|\hat R_{\lambda\mu}|\nu>$, the mixing coefficients $c_n$ and
$\bar c_n$ being determined by the structure of the wave packet
(\ref{Psi}). Naturally, the same is true for the variables
$\L_{\lambda\mu}$, $\P_{\lambda\mu}$. The dynamical equation for
$<0|\hat R_{\lambda\mu}|\nu>$ is in turn, the linear combination of
RPA equations (\ref{DD}) (or (\ref{RPA}) in the case of arbitrary
interaction) for the amplitudes $X_{pq}, Y_{pq}$, the mixing
coefficients being particle--hole matrix elements of the operator
$\hat R_{\lambda\mu}$.

As we see, there exist exact relations between the dynamical
equations for the variables of the WFM method (moments) and the RPA
dynamical equations for the amplitudes $X_{pq}, Y_{pq}$.
One should note however, that these relations are exact only in our
simplified model, because in general both methods, to be exact, have
to operate within an infinite number of dynamical equations.
In RPA one replaces the infinite number of particle--hole pairs of the
shell model by the infinite number of
phonons with the hope that the essential part of physics is described
by the small number of the lowest energy collective phonons and,
consequently, one can neglect the rest of (infinite number) phonons.
The coupling of the dynamical equations for $X_{pq}, Y_{pq}$,
corresponding to different particle--hole pairs is realized by the
matrix elements of the nucleon--nucleon interaction (see equations
(\ref{RPA})). An analogous situation is observed in the WFM method,
where the dynamical equations for Cartesian tensors of
rank $n=2$ are coupled (by the interaction terms in
(\ref{sincyclic})) with dynamical equations for tensors of rank
$n=3$, these equations being coupled with the ones for tensors of
rank $n=4$ and so on up to $n=\infty$. And again one hopes that the
essential part of physics is described by a few number of the lowest
ranks tensors. This hope is based on the evident consideration that
the higher rank tensors (moments) are responsible for more refined
detailes and that, by neglecting them, one does not appreciably
influence the description of the more global physics which is
described with the lower ranks tensors. In this game of including
only the lowest rank tensors one has to remember the trivial (but
important) rule: the highest rank of tensors must not be less than
the multipolarity of the studied motion.

It is easy to see that the nature of truncation in the two methods
is quite different. So that in practical calculations with realistic
Hamiltonians one can not establish the exact relation between these
methods unless one works in the full space in both methods.

Of course there are exceptions like the case of the mean field
potentials with quadratic coordinate dependence (harmonic oscillator
with quadrupole--quadrupole or monopole--monopole residual
interaction).
Due to the huge degeneracy of the particle--hole configuration space
all RPA sums are calculated analytically without any approximations.
The same happens in WFM method -- the dynamical equations for tensors
of different ranks decouple and one obtains a finite set of equations,
which can be solved exactly.
 As a consequence, both methods give identical results
for integral characteristics of the collective motion, such as
energies and transition probabilities.

A difference appears in the description of various distributions in
coordinate space, for example, transition densities and currents,
where the WFM method can not give the exact result, because it
deals only with integrals over the whole phase space $\{\bp,\br\}$.
However, in principle the WFM method can give any number of moments
required to construct approximate expressions for these distributions
(see below).

\subsection{Flows}

We want to know the trajectories of infinitesimal displacements of
neutrons and protons during their vibrational motion (the lines of
currents). The infinitesimal displacements are determined by the
magnitudes and directions of the nucleon velocities $\bu(\br,t)$,
given by
\begin{eqnarray}
m \rho(\br,t) \bu(\br,t)&=&
\int\! \frac{4d^3p}{ (2\pi\hbar)^3}\, \bp f(\br,\bp,t)
\nonumber\\
&=&\frac{4}{(2\pi\hbar)^{3}}
\int\! d^3s\int\! d^3p
\: \bp \exp(- i\bp\cdot
 \bs/\hbar)\rho(\br+\frac{\bs}{2}, \br-\frac{\bs}{2},t)
\nonumber\\
&=&-2i\hbar \{(\nabla-\nabla')
\rho(\br, \br',t)\}_{r=r'}
=-\frac{i\hbar}{2}\sum_{\sigma,\tau} \{(\nabla-\nabla')
\rho(\br\sigma\tau, \br'\sigma\tau,t)\}_{r=r'}
\nonumber\\
&=&-\frac{i\hbar}{2} \sum_{pq}\sum_{\sigma,\tau}
\{\phi_p^*(\br\sigma\tau)\nabla\phi_q(\br\sigma\tau)
 -\phi_q(\br\sigma\tau)\nabla\phi_p^*(\br\sigma\tau)\}
<\Psi|a_p^{\dagger}a_q|\Psi>
\nonumber\\
&=&m\sum_{pq}j_{pq}(\br)\rho_{qp}(t)
=m<\Psi|\sum_{pq}j_{pq}(\br)a_p^{\dagger}a_q|\Psi>
\nonumber\\
&=&m<\Psi|\hat J(\br)|\Psi>.
\label{velos}
\end{eqnarray}
The current density operator $\hat J(\br)$ has the standard quantum
mechanical definition \cite{Ring}:
$$\hat J(\br)
=\sum_{s=1}^A\hat j_s(\br)
=\frac{\hbar}{2mi}\sum_{s=1}^A
[\delta(\br-\hat\br_s)\nabla_s
+\nabla_s\delta(\br-\hat\br_s)]
=\sum_{pq}j_{pq}(\br)a^{\dagger}_p a_q,$$
$$j_{pq}(\br)=
<p|\frac{\hbar}{2mi}[\delta(\br-\hat\br)\nabla
+\nabla\delta(\br-\hat\br)]|q>$$
$$=\sum_{\sigma,\tau}
\frac{\hbar}{2mi}[\phi^*_p(\br\sigma\tau)\nabla\phi_q(\br\sigma\tau)-
\phi_q(\br\sigma\tau)\nabla\phi^*_p(\br\sigma\tau)]
=4\frac{\hbar}{2mi}[\phi^*_p(\br)\nabla\phi_q(\br)-
\phi_q(\br)\nabla\phi^*_p(\br)].$$
The variation of $\bu$ generated by the external field (\ref{Exfield})
is
\begin{eqnarray}
\rho^{eq}(\br)\delta \bu(\br,t)
&=&\sum_{pq}j_{pq}(\br)\rho_{qp}^{(1)}(t)
\nonumber\\
&=&\sum_{\nu}[<0|\hat J(\br)|\nu>c_{\nu}
-<\nu|\hat J(\br)|0>
\bar c_{\nu}]e^{-i\Omega t}
\nonumber\\
&&+\sum_{\nu}[<\nu|\hat J(\br)|0>c_{\nu}^*
-<0|\hat J(\br)|\nu>\bar c_{\nu}^*]e^{i\Omega t}.
\label{deltau}
\end{eqnarray}

To proceed further two ways are possible.

The first, so to say direct way, is obvious. Having solutions
(\ref{XY}) for $X_{mi}^{\nu},\,Y_{mi}^{\nu}$ we can calculate
transition currents with the help of formula
(\ref{matelem}):
\begin{eqnarray}
<0|\hat J(\br)|\nu>
=\sum_{mi}(j_{im}X_{mi}^{\nu}
+j_{mi}Y_{mi}^{\nu})
=K_{\nu}\sum_{mi}\left\{
\frac{j_{im}D_{im}^*}{E_{\nu}-\epsilon_{mi}}-
\frac{j_{mi}D_{mi}^*}{E_{\nu}+\epsilon_{mi}}
\right\}
\nonumber\\
=K_{\nu}\left\{
\sum_{mi(\Delta N=0)}\left[
\frac{j_{im}D_{im}^*}{E_{\nu}-\epsilon_{0}}-
\frac{j_{mi}D_{mi}^*}{E_{\nu}+\epsilon_{0}}
\right]+
\sum_{mi(\Delta N=2)}\left[
\frac{j_{im}D_{im}^*}{E_{\nu}-\epsilon_{2}}-
\frac{j_{mi}D_{mi}^*}{E_{\nu}+\epsilon_{2}}
\right]
\right\}.
\label{trcurr}
\end{eqnarray}
The operator $D$ has a finite number of particle--hole matrix elements
$D_{mi}$, so, in principle, the sums in (\ref{trcurr}) can be
calculated exactly. The same is true for the coefficients $c_{\nu}$.
Therefore, one can find the exact (in the frame
of RPA) result for the velocity distribution $\delta \bu(\br,t)$.
However, even in this simple model one can not find a compact
analytical expression for sums in (\ref{trcurr}) -- the field of
velocities can be constructed only numerically.

The second way allows one to derive an approximate analytical
expression for $\delta \bu(\br,t)$. The main idea lies in the
parametrization of the infinitesimal displacements ${\bf \xi}(\br,t)$
\cite{Bal}. Let us recall the main points. By definition
$\di\delta \bu_i(\br,t)=\frac{\partial \xi_i(\br,t)}{\partial t}$.
The displacement $\xi_i$ is parametrized \cite{BaSc2} by the expansion
\begin{equation}
\xi_i(\br,t)=G_i(t)+\sum_{j=1}^3G_{i,j}(t)x_j+\sum_{j,k=1}^3
G_{i,jk}(t)x_jx_k+\sum_{j,k,l=1}^3G_{i,jkl}(t)x_jx_kx_l+\cdots
\label{displ}
\end{equation}
which, in principle, is infinite, however one makes the approximation
keeping only the first terms and neglecting all the rest of
it. For example, in \cite{BaSc2} only the two first terms were
kept. It turned out, that $G_i=0$ due to the triplanar symmetry
of considered nuclei. The coefficients $G_{i,j}$ were expressed
analytically in terms of the variables $\R_{21}(t)$ and $\L_{11}(t)$.
Using the dynamical relations between $\R_{21}(t)$ and $\L_{11}(t)$
given by the last equation of the set (\ref{scis}), the final
formulae for $\xi_i(\br,t)$ were found to be
$$\xi_1=\sqrt2 B\J_{13}x_3,\quad
\xi_2=\sqrt2 B\J_{23}x_3,\quad
\xi_3=\sqrt2 A(\J_{13}x_1+\J_{23}x_2)$$
with
$$\J_{13}=(\R_{2-1}-\R_{21})/2,\quad
\J_{23}=i(\R_{2-1}+\R_{21})/2,$$
\begin{eqnarray}
A=\frac{3}{\sqrt2}[1-2\frac{\bar\omega^2}{\Omega^2}(1-\alpha)\delta]
/[Q_{00}(1-\frac{2}{3}\delta)],
\nonumber\\
B=\frac{3}{\sqrt2}[1+2\frac{\bar\omega^2}{\Omega^2}(1-\alpha)\delta]
/[Q_{00}(1+\frac{4}{3}\delta)],
\label{AiB}
\end{eqnarray}
The pole structure of the right hand side of equation (\ref{deltau})
tells us, that the transition current can be calculated by means of
an expression analogous to (\ref{Fmatel}):
\begin{equation}
<0|\hat J_i(\br)|\nu>=
\hbar\lim_{\Omega\to\Omega_{\nu}}(\Omega-\Omega_{\nu})\rho^{eq}(\br)
\overline{\dot\xi_i(\br,t)\exp(i\Omega t)}/<\nu|\hat W|0>.
\label{trandis}
\end{equation}
For the  $\xi_i$ from above we obtain (using formulae
(\ref{deltaR}) and (\ref{Cnu}))
$$<0|\hat J_1(\br)|\nu>=-i\Omega_{\nu}\rho^{eq}(\br)\frac{B}{\sqrt2}
<0|\hat R_{2-1}-\hat R_{21}|\nu>x_3,$$
$$<0|\hat J_2(\br)|\nu>=\Omega_{\nu}\rho^{eq}(\br)\frac{B}{\sqrt2}
<0|\hat R_{2-1}+\hat R_{21}|\nu>x_3,$$
$$<0|\hat J_3(\br)|\nu>=-i\Omega_{\nu}\rho^{eq}(\br)\frac{A}{\sqrt2}
[<0|\hat R_{2-1}-\hat R_{21}|\nu>x_1
+i<0|\hat R_{2-1}+\hat R_{21}|\nu>x_2].$$

It is obvious that the second way is more adequate for the WFM
method, because the moments $\R_{21}(t)$ and $\L_{11}(t)$ are just
WFM variables and the dynamical relation between them is just given
by the WFM dynamical equation.

If necessary, one can find the next term of the series (\ref{displ}).
To calculate the respective coefficients $G_{i,jkl}(t)$ in the WFM
method
one is obliged to derive (and solve) the set of dynamical equations
for higher (fourth) order moments of the Wigner function. To solve
the same problem with RPA, it is necessary to construct the analogous
set of dynamical equations for transition matrix elements of the
respective operators. The required
work is approximately of the same order of complexity in both cases.

In conclusion in full RPA one must calculate the currents numerically
leading to fine detailes (shell effects) whereas in WFM and
approximate RPA treatment one obtains their gross structure with
analytical formulas. The latter feature is quite important in order
to understand the real character of the motion under study since
current patterns produced numerically from complicated formulas with
a lot of summations like in (\ref{trcurr}) can hardly be interpreted
physically. A good example is the interplay of the scissors mode and
the isovector giant quadrupole resonance. Looking only at the flow
patterns (see Figs. 1, 2 in \cite{BaSc2}) one would not be able to
tell that the former is mostly rotational with a small amount of an
irrotational component and the other way round for the latter, as
this can be seen from eqs. (42)--(47) in \cite{BaSc2}. In this
respect it is important to work with the
infinitesimal displacements $\xi_i$, because by definition they are
differentials ($\xi_1=dx, \xi_2=dy, \xi_3=dz$) which allow one to
construct differential equations for the current fields.  For example
equation (\ref{trandis}), showing that transition current is
proportional to a differential, allows one to derive a differential
equation for the current field in RPA directly from (\ref{trcurr}).
For example
$$\frac{<0|\hat J_1(\br)|\nu>}{<0|\hat J_2(\br)|\nu>}=\frac{dx}{dy}.$$

\section{Conclusion}

The properties of collective excitations (the scissors mode, isovector
and isoscalar giant quadrupole resonances) of the harmonic oscillator
Hamiltonian with the quadrupole--quadrupole residual interaction (HO+
QQ) have been studied with two methods: WFM and RPA. We have found
that both methods give the same analytical expressions for energies
and transition probabilities of all considered excitations. This,
however, does not mean that WFM and RPA are identical approaches in
all respects. For example current distributions are described
differently in the two approaches even in this simple model.
In general both methods are not equivalent unless one makes sure that
the space of moments corresponds exactly to the particle--hole space
used in RPA. However, the spirit of WMF is rather to drastically reduce
the dimensions in considering only low order rank tensors. In this
way, of course, one will loose the fine structure in the spectrum but
still the gross structure will be well approximated. One also may check
the convergence of the method in increasing the number of moments. In the
case of well defined resonances only some more satellites to the main
peak should appear. Such a method may be particularly useful in the case
of deformed nuclei where the dimension of the RPA matrices becomes easily
prohibitive.

It makes no sense to speak about advantages or disadvantages of one
of the two discussed methods -- they are complementary. Of course, RPA
gives complete, exhaustive information concerning the microscopic
(particle--hole) structure of collective excitations. However,
sometimes considerable additional effort is required to understand
their physical nature. On the contrary, WFM method gives direct
information on the physical nature of the excitations. Our results
serve as a very good illustration of this situation. What do we learn
about the scissors mode and IVGQR from each of the two methods? RPA
says that the scissors mode is mostly created by $\Delta N=0$
particle--hole excitations with a small admixture of $\Delta N=2$
particle--hole excitations and vice versa for IVGQR. Without further
effort -- this is about all. One does even not suspect the key role
of the relative angular momentum in the creation of the scissors mode.
On the other hand, the WFM method directly reveals
that the scissors mode appears due to
oscillations of the relative angular momentum with a small admixture
of the quadrupole mode and vice versa for IVGQR.
Further, it informs us about the extremely important role of the
Fermi surface deformation in the formation of the scissors mode.
This demonstrates very well the difference between two approaches:
the RPA describes the fine structure of collective excitations
whereas the WFM method gives their gross structure.

Two new mathematical results are obtained for the HO+QQ model.
We have proved exactly, without any approximations, the orthogonality
of the "spurious" state to all physical states. In this sense, we
have generalized the result of Lo Iudice \cite{Lo96} derived in a
small deformation approximation. The analytical expressions are
derived for the normalization factor of the synthetic scissors
state and overlaps of this state with eigenstates of the model.
It is shown, that the overlap of the synthetic scissors with the
real scissors reaches its maximal value $\sim 0.6$ in a small
deformation limit. The increasing of  $\delta$ and /or taking into
account pairing correlations decreases the overlap, that is
confirmed by numerous microscopic calculations with various forces
\cite{Zaw}. Such small overlap leads inevitably to the conclusion,
that the original model of counter rotating rigid rotors \cite{Lo2000}
has not very much in common with the real scissors mode, the correct
description of which requires the proper treatment of the Fermi surface
deformation and the coupling with IVGQR.

\section*{Appendix A}

It is known that the deformed harmonic oscillator Hamiltonian can be
obtained in a Hartree approximation "by making the assumption that the
isoscalar part of the QQ force builds the one-body container well"
\cite{Hilt92}. In our case it is obtained quite easily by summing
the expressions for $V^{\rm p}$ and $V^{\rm n}$
(formula (\ref{poten})):
$$V(\br,t)=\frac{1}{2}(V^{\rm p}(\br,t)+V^{\rm n}(\br,t))
=\frac{1}{2}m\,\omega^2r^2+
\kappa_0\sum_{\mu=-2}^{2}(-1)^{\mu}
 Q_{2-\mu}(t)q_{2\mu}(\br).                     \eqno ({\rm A.}1)$$
In the state of equilibrium (i.e., in the absence of an external
field) $Q_{2\pm1}=Q_{2\pm2}=0$. Using the definition \cite{BM}
$Q_{20}=Q_{00}\frac{4}{3}\delta$ and the formula
$q_{20}=2z^2-x^2-y^2$ we obtain the potential of the
anisotropic harmonic oscillator
$$V(\br)=\frac{m}{2}[\omega_x^2(x^2+y^2)+\omega_z^2z^2]$$
with oscillator frequencies
$$\omega_x^2=\omega_y^2=\omega^2(1+\sigma\delta), \quad
\omega_z^2=\omega^2(1-2\sigma\delta),$$
where $\di \sigma=-\kappa_0\frac{8Q_{00}}{3m\omega^2}$. The
definition of the deformation parameter $\delta$ must be reproduced
by the harmonic oscillator wave functions, which allows one to fix
the value of $\sigma$. We have
$$Q_{00}=\frac{\hbar}{m}(\frac{\Sigma_x}{\omega_x}
+\frac{\Sigma_y}{\omega_y}+\frac{\Sigma_z}{\omega_z}),\quad
Q_{20}=2\frac{\hbar}{m}(\frac{\Sigma_z}{\omega_z}
-\frac{\Sigma_x}{\omega_x}),$$
where $\di \Sigma_x=\Sigma_{i=1}^A(n_x+\frac{1}{2})_i$ and $n_x$
is the oscillator quantum number.
Using the self-consistency condition \cite{BM}
$$\Sigma_x\omega_x=\Sigma_y\omega_y=\Sigma_z\omega_z=\Sigma_0\omega_0,$$
where $\Sigma_0$ and $\omega_0$ are defined in the
spherical case, we get
$$\frac{Q_{20}}{Q_{00}}=2\frac{\omega_x^2-\omega_z^2}
{\omega_x^2+2\omega_z^2}=\frac{2\sigma\delta}{1-\sigma\delta}
=\frac{4}{3}\delta.$$
Solving the last equation with respect to $\sigma$, we find
$$\sigma=\frac{2}{3+2\delta}.                \eqno ({\rm A.}2)$$
Therefore, the oscillator frequences and the strength constant can
be written as
$$\omega_x^2=\omega_y^2=\bar\omega^2(1+\frac{4}{3}\delta), \quad
\omega_z^2=\bar\omega^2(1-\frac{2}{3}\delta), \quad
\kappa_0=-\frac{m\bar\omega^2}{4Q_{00}}      \eqno ({\rm A.}3)$$
with $\bar\omega^2=\omega^2/(1+\frac{2}{3}\delta).$
The condition for volume conservation
$\omega_x\omega_y\omega_z=const=\omega_0^3$ makes $\omega$
$\delta$-dependent
$$\omega^2=\omega_0^2\frac{1+\frac{2}{3}\delta}
{(1+\frac{4}{3}\delta)^{2/3}(1-\frac{2}{3}\delta)^{1/3}}
.$$
So the final expressions for oscillator frequences are
$$\omega_x^2=\omega_y^2=\omega_0^2\left(\frac{1+\frac{4}{3}\delta}
{1-\frac{2}{3}\delta}\right)^{1/3}, \quad
\omega_z^2=\omega_0^2\left(\frac{1-\frac{2}{3}\delta}
{1+\frac{4}{3}\delta}\right)^{2/3}.           \eqno ({\rm A.}4)$$

It is interesting to compare these expressions with the very popular
\cite{BM,Ring} parametrization
$$\omega_x^2=\omega_y^2=\omega'^2(1+\frac{2}{3}\delta'), \quad
\omega_z^2=\omega'^2(1-\frac{4}{3}\delta').$$
The volume conservation condition gives
$$\omega'^2=\frac{\omega_0^2}
{(1+\frac{2}{3}\delta')^{2/3}(1-\frac{4}{3}\delta')^{1/3}},$$
so the final expressions for oscillator frequences are
$$\omega_x^2=\omega_y^2=\omega_0^2\left(\frac{1+\frac{2}{3}\delta'}
{1-\frac{4}{3}\delta'}\right)^{1/3}, \quad
\omega_z^2=\omega_0^2\left(\frac{1-\frac{4}{3}\delta'}
{1+\frac{2}{3}\delta'}\right)^{2/3}.          \eqno ({\rm A.}5)$$
The direct comparison of expressions ({\rm A.}4) and ({\rm A.}5)
allows one to esatablish the following relation between $\delta$ and
$\delta'$:
$$\delta'=\frac{\delta}{1+2\delta},\quad
\delta=\frac{\delta'}{1-2\delta'}.$$
One more parametrization of oscillator frequences can be
found in the review \cite{Zaw}:
$$\omega_x^2=\omega_y^2=\frac{\omega"^2}{1-\frac{2}{3}\delta"}, \quad
\omega_z^2=\frac{\omega"^2}{1+\frac{4}{3}\delta"}.$$
One has from the volume conservation condition
$$\omega"^2=\omega_0^2(1-\frac{2}{3}\delta")^{2/3}
(1+\frac{4}{3}\delta")^{1/3},$$
so the final expressions for oscillator frequences are
$$\omega_x^2=\omega_y^2=\omega_0^2\left(\frac{1+\frac{4}{3}\delta"}
{1-\frac{2}{3}\delta"}\right)^{1/3}, \quad
\omega_z^2=\omega_0^2\left(\frac{1-\frac{2}{3}\delta"}
{1+\frac{4}{3}\delta"}\right)^{2/3},          \eqno ({\rm A.}6)$$
that coincide exactly with ({\rm A.}4), i.e.
$\delta"=\delta$.

It is easy to see that equations ({\rm A.}4) correspond to the
case when the
deformed density $\rho(\br)$ is obtained from the spherical density
$\rho_0(r)$ by the scale transformation \cite{Suzuki}
$$(x,y,z)\rightarrow (xe^{\alpha/2},ye^{\alpha/2},ze^{-\alpha})$$
with
$$e^{\alpha}=\left(\frac{1+\frac{4}{3}\delta}
{1-\frac{2}{3}\delta}\right)^{1/3},\quad \delta=\frac{3}{2}
\frac{e^{3\alpha}-1}{e^{3\alpha}+2},          \eqno ({\rm A.}7)$$
which conserves the volume and does not destroy the self-consistency,
because the density and potential are transformed in the same way.

It is necessary to note that $Q_{00}$ also depends on $\delta$
$$Q_{00}=\frac{\hbar}{m}(\frac{\Sigma_x}{\omega_x}
+\frac{\Sigma_y}{\omega_y}+\frac{\Sigma_z}{\omega_z})=
\frac{\hbar}{m}\Sigma_0\omega_0
(\frac{2}{\omega_x^2}+\frac{1}{\omega_z^2})=
Q_{00}^0\frac{1}
{(1+\frac{4}{3}\delta)^{1/3}(1-\frac{2}{3}\delta)^{2/3}},$$
where $Q_{00}^0=A\frac{3}{5}R^2,\,R=r_0A^{1/3}.$
As a result, the final expression for the strength constant becomes
$$\kappa_0=-\frac{m\omega_0^2}{4Q_{00}^0}
\left(\frac{1-\frac{2}{3}\delta}
{1+\frac{4}{3}\delta}\right)^{1/3}
=-\frac{m\omega_0^2}{4Q_{00}^0}e^{-\alpha},$$
that coincides with the respective result of \cite{Suzuki}.

\section*{Appendix B}

To calculate the sums
 $\di D_0=\sum_{mi(\Delta N=0)}|D_{mi}|^2$ and
$\di D_2=\sum_{mi(\Delta N=2)}|D_{mi}|^2$
 we employ the sum-rule techniques
of Suzuki and Rowe \cite{Suzuki}.
The well known harmonic oscillator
relations
$$x\psi_{n_x}=\sqrt{\frac{\hbar}{2m\omega_x}}
(\sqrt{n_x}\psi_{n_x-1}+\sqrt{n_x+1}\psi_{n_x+1}),$$
$$\hat p_x\psi_{n_x}=-i\sqrt{\frac{m\hbar\omega_x}{2}}
(\sqrt{n_x}\psi_{n_x-1}-\sqrt{n_x+1}\psi_{n_x+1}) \eqno ({\rm B.}1)$$
allow us to write
$$xz\psi_{n_x}\psi_{n_z}=\frac{\hbar}{2m\sqrt{\omega_x\omega_z}}
(\sqrt{n_xn_z}\psi_{n_x-1}\psi_{n_z-1}
+\sqrt{(n_x+1)(n_z+1)}\psi_{n_x+1}\psi_{n_z+1}$$
$$+\sqrt{(n_x+1)n_z}\psi_{n_x+1}\psi_{n_z-1}
+\sqrt{n_x(n_z+1)}\psi_{n_x-1}\psi_{n_z+1}),$$
$$\frac{\hat p_x\hat p_z}{m^2\omega_x\omega_z}
\psi_{n_x}\psi_{n_z}
=-\frac{\hbar}{2m\sqrt{\omega_x\omega_z}}
(\sqrt{n_xn_z}\psi_{n_x-1}\psi_{n_z-1}
+\sqrt{(n_x+1)(n_z+1)}\psi_{n_x+1}\psi_{n_z+1}$$
$$-\sqrt{(n_x+1)n_z}\psi_{n_x+1}\psi_{n_z-1}
-\sqrt{n_x(n_z+1)}\psi_{n_x-1}\psi_{n_z+1}).     \eqno ({\rm B.}2)$$
These formulae demonstrate in an obvious way that the operators
$$P_0=\frac{1}{2}(zx+
\frac{1}{m^2\omega_x\omega_z}\hat p_x\hat p_z)
\quad\mbox{and}\quad
P_2=\frac{1}{2}(zx-
\frac{1}{m^2\omega_x\omega_z}\hat p_x\hat p_z)$$
contribute only to the excitation of the
$\Delta N=0$ and $\Delta N=2$ states, respectively.
Following \cite{Suzuki}, we express the $zx$ component of
$r^2Y_{21}=\sqrt{\frac{5}{16\pi}}D
=-\sqrt{\frac{15}{8\pi}}z(x+iy)$ as
$$zx=P_0+P_2.$$

 Hence, we have
$$\epsilon_0\sum_{mi(\Delta N=0)}|<0|\sum_{s=1}^Az_sx_s|mi>|^2
=\epsilon_0\sum_{mi}|<0|\sum_{s=1}^AP_0(s)|mi>|^2$$
$$=\frac{1}{2}<0|[\sum_{s=1}^A P_0(s),[H,\sum_{s=1}^A P_0(s)]]|0>,
						 \eqno ({\rm B.}3)$$
where $\epsilon_0=\hbar(\omega_x-\omega_z)$.
 The above commutator is easily evaluated for the Hamiltonian with
the potential ({\rm A.}1), as
$$<0|[\sum_{s=1}^A P_0(s),[H,\sum_{s=1}^A P_0(s)]]|0>=
\frac{\hbar}{2m}\epsilon_0
\left(\frac{<0|\sum_{s=1}^Az_s^2|0>}{\omega_x}
-\frac{<0|\sum_{s=1}^Ax_s^2|0>}{\omega_z}\right).\eqno ({\rm B.}4)$$
Taking into account the axial symmetry and using the definitions
$$Q_{00}=<0|\sum_{s=1}^A(2x_s^2+z_s^2)|0>,\quad
  Q_{20}=2<0|\sum_{s=1}^A(z_s^2-x_s^2)|0>,\quad
Q_{20}=Q_{00}\frac{4}{3}\delta,$$
we transform this expression to
$$<0|[\sum_{s=1}^A P_0(s),[H,\sum_{s=1}^A P_0(s)]]|0>=
\frac{\hbar}{6m}\epsilon_0 Q_{00}
\left(\frac{1+\frac{4}{3}\delta}{\omega_x}
-\frac{1-\frac{2}{3}\delta}{\omega_z}\right).    \eqno ({\rm B.}5)$$
With the help of the self-consistent expressions for
$\omega_x,\,\omega_z$ ({\rm A.}3) one comes to the following result:
$$<0|[\sum_{s=1}^A P_0(s),[H,\sum_{s=1}^A P_0(s)]]|0>=
\frac{Q_{00}}{6m}
\frac{\epsilon_0^2}{\bar\omega^2}=
\frac{\hbar^2}{6m}Q_{00}^0
\left(\frac{\omega_0}{\omega_z}
-\frac{\omega_0}{\omega_x}\right)^2.             \eqno ({\rm B.}6)$$
 By using the fact that the matrix elements for the $zy$ component of
$r^2Y_{21}$ are identical to those for the $zx$ component, because of
axial symmetry, we finally obtain
$$\epsilon_0\sum_{mi(\Delta N=0)}|<0|\sum_{s=1}^Ar^2_sY_{21}|mi>|^2=
\frac{5}{16\pi}\frac{Q_{00}}{m\bar\omega^2}\epsilon_0^2=
\frac{5}{16\pi}\frac{Q_{00}^0}{m}\frac{\epsilon_0^2}{\omega_0^2}
\left(\frac{1+\frac{4}{3}\delta}{1-\frac{2}{3}\delta}\right)^{1/3}.
						 \eqno ({\rm B.}7)$$
By calculating a double commutator for the
$P_2$ operator, we find
$$\epsilon_2\sum_{mi(\Delta N=2)}|<0|\sum_{s=1}^Ar^2_sY_{21}|mi>|^2=
\frac{5}{16\pi}\frac{Q_{00}}{m\bar\omega^2}\epsilon_2^2=
\frac{5}{16\pi}\frac{Q_{00}^0}{m}\frac{\epsilon_2^2}{\omega_0^2}
\left(\frac{1+\frac{4}{3}\delta}{1-\frac{2}{3}\delta}\right)^{1/3},
						 \eqno ({\rm B.}8)$$
where $\epsilon_2=\hbar(\omega_x+\omega_z)$.

We need also the sums $D_0^{\tau}$ and $D_2^{\tau}$ calculated
separately for neutron and proton systems with the mean fields
$V^{\rm n}$ and $V^{\rm p}$, respectively. The necessary formulae
are easily derivable from the already obtained results.
There are no any reasons to require the fulfillment of the
self-consistency conditions for neutrons and protons separately,
so one has to use formula ({\rm B.}5). The trivial change of
notation gives
$$<0|[\sum_{s=1}^Z P_0(s),[H^{\rm p},\sum_{s=1}^Z P_0(s)]]|0>=
\frac{\hbar}{6m}\epsilon_0^{\rm p} Q_{00}^{\rm p}
\left(\frac{1+\frac{4}{3}\delta^{\rm p}}{\omega_x^{\rm p}}
-\frac{1-\frac{2}{3}\delta^{\rm p}}{\omega_z^{\rm p}}\right),
						 \eqno ({\rm B.}9)$$

$$\epsilon_0^{\rm p}\sum_{mi(\Delta N=0)}
|<0|\sum_{s=1}^Zr^2_sY_{21}|mi>|^2=
\frac{5}{16\pi}\frac{\hbar}{m}\epsilon_0^{\rm p} Q_{00}^{\rm p}
\left(\frac{1+\frac{4}{3}\delta^{\rm p}}{\omega_x^{\rm p}}
-\frac{1-\frac{2}{3}\delta^{\rm p}}{\omega_z^{\rm p}}\right),
						 \eqno ({\rm B.}10)$$

$$\epsilon_2^{\rm p}\sum_{mi(\Delta N=2)}
|<0|\sum_{s=1}^Zr^2_sY_{21}|mi>|^2=
\frac{5}{16\pi}\frac{\hbar}{m}\epsilon_2^{\rm p} Q_{00}^{\rm p}
\left(\frac{1+\frac{4}{3}\delta^{\rm p}}{\omega_x^{\rm p}}
+\frac{1-\frac{2}{3}\delta^{\rm p}}{\omega_z^{\rm p}}\right).
						 \eqno ({\rm B.}11)$$
The nontrivial information is contained in oscillator frequences of
the mean fields $V^{\rm p}$ and $V^{\rm n}$ (formula (\ref{poten}))
$$(\omega_x^{\rm p})^2=\omega^2[1-\frac{2}{m\omega^2}
(\kappa Q_{20}^{\rm p}+\bar\kappa Q_{20}^{\rm n})],\quad
(\omega_z^{\rm p})^2=\omega^2[1+\frac{4}{m\omega^2}
(\kappa Q_{20}^{\rm p}+\bar\kappa Q_{20}^{\rm n})],$$
$$(\omega_x^{\rm n})^2=\omega^2[1-\frac{2}{m\omega^2}
(\kappa Q_{20}^{\rm n}+\bar\kappa Q_{20}^{\rm p})],\quad
(\omega_z^{\rm n})^2=\omega^2[1+\frac{4}{m\omega^2}
(\kappa Q_{20}^{\rm n}+\bar\kappa Q_{20}^{\rm p})].
						 \eqno ({\rm B.}12)$$

The above-written formulae can be used also to calculate the analogous
sums for operators containing various combinations of momenta and
coordinates, for example, components of an angular momentum,
tensor products $(r\hat p)_{21}$ and $(\hat p^2)_{21}$.
Really, by definition $\hat I_1=y\hat p_z-z\hat p_y, \quad
\hat I_2=z\hat p_x-x\hat p_z$.
 In accordance with ({\rm B.}1), we have
$$x\hat p_z\psi_{n_x}\psi_{n_z}=-i\frac{\hbar}{2}
\sqrt{\frac{\omega_z}{\omega_x}}
(\sqrt{n_xn_z}\psi_{n_x-1}\psi_{n_z-1}
-\sqrt{(n_x+1)(n_z+1)}\psi_{n_x+1}\psi_{n_z+1}$$
$$+\sqrt{(n_x+1)n_z}\psi_{n_x+1}\psi_{n_z-1}
-\sqrt{n_x(n_z+1)}\psi_{n_x-1}\psi_{n_z+1}).     \eqno ({\rm B.}13)$$
 Therefore,
$$\hat I_2\psi_{n_x}\psi_{n_z}=
i\frac{\hbar}{2}(\sqrt{\frac{\omega_z}{\omega_x}}
-\sqrt{\frac{\omega_x}{\omega_z}})
(\sqrt{n_xn_z}\psi_{n_x-1}\psi_{n_z-1}
-\sqrt{(n_x+1)(n_z+1)}\psi_{n_x+1}\psi_{n_z+1})$$
$$+i\frac{\hbar}{2}(\sqrt{\frac{\omega_z}{\omega_x}}
+\sqrt{\frac{\omega_x}{\omega_z}})
(\sqrt{(n_x+1)n_z}\psi_{n_x+1}\psi_{n_z-1}
-\sqrt{n_x(n_z+1)}\psi_{n_x-1}\psi_{n_z+1}).     \eqno ({\rm B.}14)$$
Having formulae ({\rm B.}2) and ({\rm B.}14),
one derives the following expressions for
matrix elements coupling the ground state with $\Delta N=2$ and
$\Delta N=0$ excitations:
$$<n_x+1,n_z+1|\hat I_2|0>=i\frac{\hbar}{2}
\frac{(\omega_x^2-\omega_z^2)}{\omega_x+\omega_z}
\sqrt{\frac{(n_x+1)(n_z+1)}{\omega_x\omega_z}},$$
$$<n_x+1,n_z-1|\hat I_2|0>=i\frac{\hbar}{2}
\frac{(\omega_x^2-\omega_z^2)}{\omega_x-\omega_z}
\sqrt{\frac{(n_x+1)n_z}{\omega_x\omega_z}},$$
$$<n_x+1,n_z+1|xz|0>=
\frac{\hbar}{2m}\sqrt{\frac{(n_x+1)(n_z+1)}{\omega_x\omega_z}},$$
$$<n_x+1,n_z-1|xz|0>=
\frac{\hbar}{2m}\sqrt{\frac{(n_x+1)n_z}{\omega_x\omega_z}}.
						 \eqno ({\rm B.}15)$$
It is easy to see that
$$<n_x+1,n_z+1|\hat I_2|0>=
im\frac{(\omega_x^2-\omega_z^2)}{\omega_x+\omega_z}<n_x+1,n_z+1|xz|0>,$$
$$<n_x+1,n_z-1|\hat I_2|0>=
im\frac{(\omega_x^2-\omega_z^2)}{\omega_x-\omega_z}<n_x+1,n_z-1|xz|0>.$$
Due to the degeneracy of the model all particle--hole excitations with
$\Delta N=2$ have the same energy $\epsilon_2$
and all particle--hole excitations with $\Delta N=0$  have the energy
$\epsilon_0$. This fact allows one to join the
last two formulae into one general expression
$$<ph|\hat I_2|0>=i\hbar m\frac{(\omega_x^2-\omega_z^2)}
{\epsilon_{ph}}<ph|xz|0>.$$
Taking into account the axial symmetry we can write the analogous
formula for $\hat I_1$:
$$<ph|\hat I_1|0>=-i\hbar m\frac{(\omega_x^2-\omega_z^2)}
{\epsilon_{ph}}<ph|yz|0>.$$
The magnetic transition operator (\ref{Omagn}) is proportional
 to the angular momentum:
$\di\hat f_{1\pm1}=
-\frac{ie}{4mc}\sqrt{\frac{3}{2\pi}}(\hat I_2\mp i\hat I_1)$
Therefore, we can write
$$<ph|\hat f_{1\pm1}|0>=-\frac{e\hbar}{2c\sqrt5}
\frac{(\omega_x^2-\omega_z^2)}{\epsilon_{ph}}<ph|r^2Y_{2\pm1}|0>.
						 \eqno ({\rm B.}16)$$
Similar calculations for the tensor product
$(r\hat p)_{21}
=-\frac{1}{2}[z\hat p_x+x\hat p_z+i(z\hat p_y+y\hat p_z)]$
lead to the following relation:
$$<ph|(r\hat p)_{21}|0>
=i\frac{m}{\hbar}\sqrt{\frac{2\pi}{15}}
\epsilon_{ph}<ph|r^2Y_{2\pm1}|0>
=i\frac{m}{2\hbar}\epsilon_{ph}<ph|r^2_{21}|0>.  \eqno ({\rm B.}17)$$

Two kinds of particle--hole matrix elements are obtained from the
second formula of ({\rm B.}2):
      $$<n_x+1,n_z+1|\hat p_x\hat p_z|0>=-\hbar m\omega_x\omega_z
\sqrt{\frac{(n_x+1)(n_z+1)}{2\omega_x2\omega_z}},$$
$$<n_x+1,n_z-1|\hat p_x\hat p_z|0>=\hbar m\omega_x\omega_z
\sqrt{\frac{(n_x+1)n_z}{2\omega_x2\omega_z}}.$$
Simple comparison with ({\rm B.}15) shows that
      $$<n_x+1,n_z+1|\hat p_x\hat p_z|0>=
-m^2\omega_x\omega_z<n_x+1,n_z+1|xz|0>,$$
      $$<n_x+1,n_z-1|\hat p_x\hat p_z|0>=
m^2\omega_x\omega_z<n_x+1,n_z-1|xz|0>.$$
With the help of obvious relations
        $$2\omega_x\omega_z=
\omega_x^2+\omega_z^2-\epsilon_0^2/\hbar^2,\quad
-2\omega_x\omega_z=\omega_x^2+\omega_z^2-\epsilon_2^2/\hbar^2$$
these two formulae can be joined into one expression
      $$<ph|\hat p_x\hat p_z|0>=\frac{m^2}{2}
(\omega_x^2+\omega_z^2-\epsilon_{ph}^2/\hbar^2)<ph|xz|0>.$$
By definition $\hat p^2_{21}=-\hat p_z(\hat p_x+i\hat p_y)$ and
$\hat r^2_{21}=-z(x+iy)$, hence,
$$<ph|\hat p^2_{21}|0>=\frac{m^2}{2}(\omega_x^2+\omega_z^2
-\epsilon_{ph}^2/\hbar^2)<ph|r^2_{21}|0>.        \eqno ({\rm B.}18)$$

\end{document}